\documentclass[graybox]{svmult}

\usepackage{setspace}
\usepackage{listings}
\usepackage{ marvosym }
\usepackage{type1cm}        
\usepackage{float}
\usepackage{makeidx}         
\usepackage{graphicx}        
\usepackage{multicol}        
\usepackage[bottom]{footmisc}

\graphicspath{{Figure/}}
\usepackage{ amssymb }

\usepackage{newtxtext}       %
\usepackage{newtxmath}       
\usepackage{algorithm}
\usepackage{algorithmicx}
\usepackage{algpseudocode}
\usepackage{amsmath}
\usepackage{caption}
\floatname{algorithm}{Algorithm}

\renewcommand{\algorithmicensure}{\textbf{Output:}}

\definecolor{codegreen}{rgb}{0,0.6,0}
\definecolor{codegray}{rgb}{0.5,0.5,0.5}
\definecolor{codepurple}{rgb}{0.58,0,0.82}
\definecolor{backcolour}{rgb}{0.95,0.95,0.92}
\floatname{algorithm}{}

\lstdefinestyle{mystyle}{
  backgroundcolor=\color{backcolour},   commentstyle=\color{codegreen},
  keywordstyle=\color{magenta},
  numberstyle=\tiny\color{codegray},
  stringstyle=\color{codepurple},
  basicstyle=\ttfamily\footnotesize,
  breakatwhitespace=false,
  breaklines=true,
  captionpos=b,
  keepspaces=true,
  numbers=left,
  numbersep=5pt,
  showspaces=false,
  showstringspaces=false,
  showtabs=false,
  tabsize=2
}

\makeindex             

\usepackage{nameref}
\begin{document}

\title*{HexDom: Polycube-Based Hexahedral-Dominant Mesh Generation}
\titlerunning{HexDom}
\author{Yuxuan Yu, Jialei Ginny Liu and Yongjie Jessica Zhang}
\authorrunning{Y. Yu, J. G. Liu, and Y. J. Zhang}
\institute{Y.~Yu \at Department of Mechanical Engineering, Carnegie Mellon University, Pittsburgh, PA 15213, USA \email{yuxuany1@andrew.cmu.edu}  \and J. G.~Liu \at Department of Mechanical Engineering, Carnegie Mellon University, Pittsburgh, PA 15213, USA \email{jialeil@andrew.cmu.edu} \and Y. J.~ Zhang (\Letter) \at Department of Mechanical Engineering, Carnegie Mellon University, Pittsburgh, PA 15213, USA \email{jessicaz@andrew.cmu.edu}}
%
%
\maketitle

\abstract{In this paper, we extend our earlier polycube-based
  all-hexahedral mesh generation method to hexahedral-dominant mesh
  generation, and present the HexDom software package. Given the
  boundary representation of a solid model, HexDom creates a
  hex-dominant mesh by using a semi-automated polycube-based mesh
  generation method. The resulting hexahedral dominant mesh includes
  hexahedra, tetrahedra, and triangular prisms. By adding
  non-hexahedral elements, we are able to generate better quality
  hexahedral elements than in all-hexahedral meshes. We explain the
  underlying algorithms in four modules including segmentation,
  polycube construction, hex-dominant mesh generation and quality
  improvement, and use a rockerarm model to explain how to run the
  software. We also apply our software to a number of other complex
  models to test their robustness. The software package and all tested
  models are availabe in github (https://github.com/CMU-CBML/HexDom).}

\section{Introduction}

In finite element analysis (FEA), a 3D domain can be discretized into
tetrahedral or hexahedral (hex) meshes. For tetrahedral mesh
generation, various strategies have been proposed in the
literature~\cite{ref:zhangbook,khan2020surface,liang2014octree,qian2013intersection},
such as octree-based~\cite{SG91,ZhangJ2005}, delaunay
triangulation~\cite{delaunay1934sphere}, and advancing front
methods~\cite{frey1996delaunay,lohner1988three,seveno1997towards}. Because
tetrahedral meshes can be created automatically, it has been widely
used in industry. However, to achieve the same precision in FEA, a
tetrahedral mesh requires more elements than an all-hex mesh
does. As a result, many techniques have been developed to
  generate all-hex meshes~\cite{xu2021singularity,
    xie2020interpolatory,Liang2013,qian2012automatic,qian2010quality}
  or converting imaging data to all-hex meshes~\cite{ref:zhangbook,
    zhang2013challenges, ZhangJ2006}. Also, hex meshes can serve as
  multiple-material domains~\cite{ZhangJ2010,ZQ2012} or input control
  meshes for
  IGA~\cite{Lai2016,wei2018blended,wei17a,Lai2017,ref:pan14,Wenyan2013c,Wenyan2011a,wenyan2010a}. Some
applications of hex mesh generation in new engineering applications
can also be found in~\cite{yu2019anatomically, yu2019thermoplastic,
  li2019isogeometric, zhang2012atlas, Zhang20072943}. Several
literatures develop methods for unstructured hex mesh generation, such
as grid-based or octree-based \cite{schneiders1996grid, S97}, medial
surface \cite{Price1995,price1997hexahedral},
plastering~\cite{Blacker1991,SKOB06}, whisker weaving~\cite{FM99}, and
vector field-based methods \cite{nieser2011cubecover}. These methods
have been used to create hex meshes for certain geometries, but are
not robust and reliable for arbitrary geometries. On the other hand,
although an all-hex mesh provides a more accurate solution, a
high-quality all-hex mesh is more difficult to create automatically.

Compared with all-hex mesh generation, a hex-dominant mesh generation,
which combines advantages of both tetrahedral and hex elements, is
more automatic and robust for complex solid models. In the literature,
several strategies have been proposed to generate hex-dominant
meshes. An indirect method was suggested by~\cite{yamakawa2003fully},
the domain is first meshed into tetrahedral elements and then merged
into a hex-dominant mesh with the packing technology.  Several other
hex-dominant meshing techniques were also presented
in~\cite{owen2000h, meyers1998hex, meshkat2000generating}. Such
indirect methods create hex-dominant meshes with too many
singularities, and the tetrahedral mesh directly influences the
quality of the hex-dominant mesh. Similar to unstructured all-hex mesh
generation, the direct method is also more preferable for hex-dominant
meshes~\cite{TW00,Owen1998}. The polycube-based
method~\cite{Tarini2004, Gregson2011} is an attractive direct approach
to obtain hex-dominant meshes by using degenerated cubes. The
polycube-based method was mainly used for all-hex meshing. A smooth
harmonic field~\cite{LeiLiu2012a} was used to generate polycubes for
arbitrary genus geometry. Boolean operations~\cite{LZH2014} were
introduced to deal with arbitrary genus geometry. In~\cite{Liu2015}, a
polycube structure was generated based on the skeletal branches of a
geometric model. Using these methods, the structure of the polycube
and the mapping distortion greatly influence the quality of the hex
mesh. The calculation of the polycube structure with a low mapping
distortion remains an open problem for complex geometry. It is important
to improve the quality of the mesh for analysis by using methods such
as pillowing, smoothing, and optimization~\cite{qian2010quality,
  YZhang2009c, wenyan2011b, qian2012automatic}. Pillowing is an
insert-sheet technique that eliminates situations where two adjacent
hex elements share more than one face. Smoothing and optimizing are
used to further improve the quality of the mesh by moving the
vertices. In our software, we implement all of the above methods to
improve the quality of hex elements.

In this paper, we extend our earlier semi-automatic polycube-based
all-hex generation to hex-dominant meshing. The software package
includes: 1) polycube based geometric decomposition of a surface
triangle mesh; 2) generation of the polycube consisting of
non-degenerated and degenerated cubes; 3) creation of a parametric
domain for different types of degenerated unit cubes including prisms
and tetrahedra; and 4) creation of a hex-dominant mesh. We first go
through the entire pipeline and explain the algorithm behind each
module of the pipeline. Then, we use a specific example to follow all
the steps and run the software. In particular, when user intervention
is required, the details of manual work are explained. The paper is
organized as follows. In Section 2 we provide an overview of the
pipeline. In Section 3 we present the HexDom software package, with a
semi-automatic polycube-based hex-dominant mesh generateion of a CAD
file. Finally, in Section 4 we show various complex models with our
software package.

\section{Pipeline design}

\begin{figure}[!htb]
\center{\includegraphics[width=\linewidth]{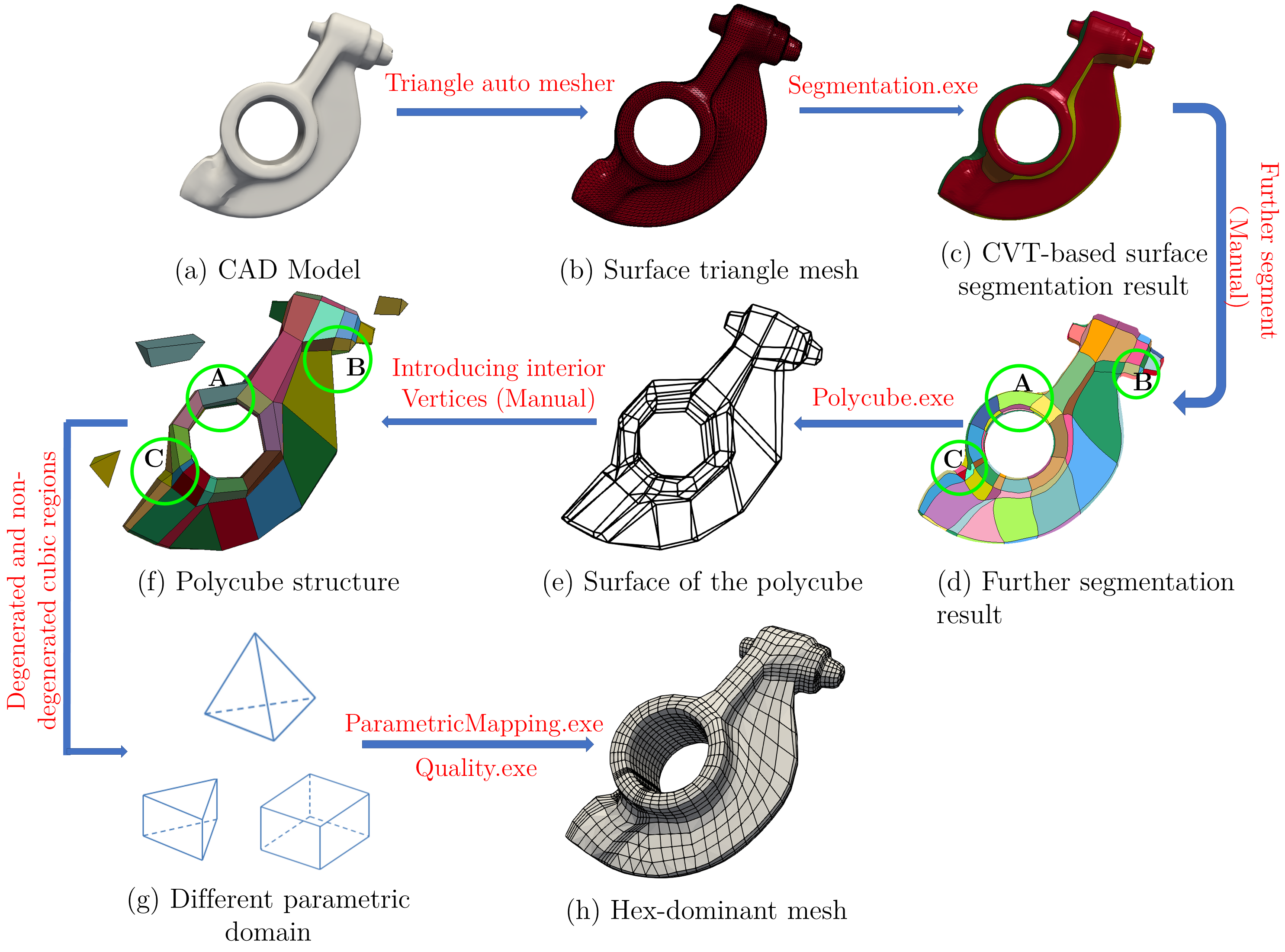}
  \caption{\label{fig:Pipeline_Structure}The HexDom software package. For each
    process, the black texts describe the object and the red texts show the
    operation needed to go to the next process. Manual work is involved in
    further segmentation and introducing interior vertices. Regions A, B and C
    (green circles) in (d, f) contain a hex, prism and tetrahedral shaped
    structure, respectively.}}
\end{figure}
Our pipeline uses polycube-based method to create a hex-dominant mesh
from an input CAD model. As shown in
Fig.~\ref{fig:Pipeline_Structure}, we first generate a triangle mesh
from the CAD model by using the free software LS-PrePost. Then we use
centroidal Voronoi tessellation (CVT)
segmentation~\cite{HZ2015CMAME,hu2018cvt,HZL2016} to create a
polycube structure~\cite{Tarini2004}. The polycube structure consists
of multiple non-degenerated cubes and degenerated cubes. The
non-degenerated cubes will yield hex elements via parametric
mapping~\cite{floater1997} and octree
subdivision~\cite{wenyan2011b}. The degenerated cubes will yield
degenerated elements such as prisms and tetrahedra in the final mesh
. Here, we implement the subdivision algorithm separately for
prism-shape regions and tetrahedral-shape regions. The quality of the
hex dominant mesh is evaluated to ensure that the resulting mesh can
be used in FEA. In case that a poor quality hex element is generated
in hex-dominant meshes, the program has various quality improvement
functions, including pillowing~\cite{YZhang2009c}, smoothing, and
optimization~\cite{qian2012automatic}. Each quality improvement
function can be performed independently and one can use these
functions to improve the mesh quality. Currently, our software only
has a command-line interface (CLI). Users need to provide the required
options on the command line to run the software. In
Section~\ref{sec:hexg-polyc-based}, we will explain in detail the
algorithms implemented in the software as well as how to run the
software.

\section{HexDom: Polycube-based hex-dominant mesh generation}
\label{sec:hexg-polyc-based}

Surface segmentation, polycube construction, parametric mapping, and
subdivision are used together in the HexDom software package to
generate a hex-dominant mesh from the boundary representation of the
input CAD model. Given a triangle mesh generated from the CAD model,
we first use surface segmentation to divide the mesh into several
surface patches that meet the restrictions of the polycube structure,
which will be discussed in Section~\ref{sec:surface-partition}. The
corner vertices, edges, and faces of each surface patch are then
extracted from the surface segmentation result to construct a polycube
structure. Each component of the polycube structure is topologically
equivalent to a cube or a degenerated cube. Finally, we generate the
hex-dominant mesh through parametric mapping and subdivision. Quality
improvement techniques can be used to further improve the mesh
quality.

In this section, we will introduce the main algorithm for each module of
the HexDom software package, namely surface segmentation,
polycube construction, parametric mapping and subdivision, and quality
improvement. We will use a rockerarm model (see
Fig.~\ref{fig:Pipeline_Structure}) to explain how to run CLI for each
module. We will also discuss the user intervention involved in the
semi-automatic polycube-based hex-dominant mesh generation.

\subsection{Surface segmentation}
\label{sec:surface-partition}

The surface segmentation in the pipeline framework is implemented
based on CVT
segmentation~\cite{HZ2015CMAME,hu2018cvt,HZL2016}. CVT
segmentation is used to classify vertices by minimizing an energy
function. Each group is called a Voronoi region and has a
corresponding center called a generator. The Voronoi region and its
corresponding generator are updated iteratively in the minimization
process. In~\cite{HZ2015CMAME}, each element of the surface triangle
mesh is assigned to one of the six Voronoi regions based on the normal
vector of the surface. The initial generators of the Voronoi regions
are the three principal normal vectors and their opposite normals
vectors ($ \pm X$, $ \pm Y$, $ \pm Z$). Two energy functions and their
corresponding distance functions are used together
in~\cite{HZ2015CMAME}. The classical energy function and its
corresponding distance function provide initial Voronoi regions and
generators. Then, the harmonic boundary-enhanced (HBE) energy function
and its corresponding distance function are applied to eliminate
non-monotone boundaries. The detailed definitions of the energy
functions and their corresponding distance functions are described in
\cite{HZ2015CMAME}. The surface segmentation process was also
summarized in the \textbf{Surface Segmentation Algorithm}
in~\cite{yu2020hexgen}.

Once we get the initial segmentation result, we need to further
segment each Voronoi region into several patches to satisfy the
topological constraints for polycube construction (see
Fig.~\ref{fig:Pipeline_Structure}(d)).  We use two types of
patches. The first type of segmented surface patch corresponds to one
boundary surface of the non-degenerated cubes and quadrilateral surface of the
prism-shape degenerated cubes. The second type of segmented surface
patch corresponds to one triangular boundary surface of the degenerated
cubes. The choice of types of patches depends on the following three
criteria: 1) geometric features such as sharp corners with small
angles and prism/tetrahedral-like features; 2) critical
regions based on finite element simulation, such as regions with the
maximum stress/strain and regions with a high load; and 3) requirements
from user applications which enhance the capability of user
interaction.  For the first type of segmented surface patch, the
following three conditions should be satisfied during the further
segmentation: 1) two patches with opposite orientations (e.g., +X and
-X) cannot share a boundary; 2) each corner vertex must be shared by
more than two patches; and 3) each patch must have four boundaries.
For the second type of segmented surface patch, we modified the third
conditions to that each patch must have three boundaries.

Note that we define the corner vertex as a vertex locating at the
corner of the cubic region or degenerated cubic region in the
model. The further segmentation is done manually by using the patch ID
reassigning function in LS-PrePost. The detailed operation was shown
in~\cite{yu2020hexgen}.


\subsection{Polycube construction}
\label{sec:poly_construction}

In this section, we discuss the detailed algorithm of polycube
construction based on the segmented triangle mesh. Several automatic
polycube construction algorithms have been proposed in the
literature~\cite{He2009369,LinJ2008,HZ2015CMAME}, but it is
challenging to apply these methods to complex CAD models. The polycube
structure does not contain degenerated cubes either. Differently, the
polycube in this paper consists of cubes and degenerated cubes and is
topologically equivalent to the original geometry. To achieve
versatility for real industrial applications, we develop a
semi-automatic polycube construction software based on the segmented
surface. However, for some complex geometries, the process may be
slower due to potentially heavy user intervention.

The most important information we need for a polycube is its corners
and the connectivity relationship among them. For the surface of
polycube, we can automatically get the corner points and build their
connectivity based on the segmentation result by using the algorithm
similar to the \textbf{Polycube Boundary Surface Construction
  Algorithm} in~\cite{yu2020hexgen}. The difference is that we need to
adjust the implementation based on different patch types: finding its
three corners for a triangular patch and finding its four corners for
a quadrilateral patch. It is difficult to obtain inner vertices and
their connectivity because we only have a surface input with no
information about the interior volume. In fact, this is where users
need to intervene. We use LS-PrePost to manually build the interior
connectivity. You can find the detailed operation in Appendix A3
in~\cite{yu2020hexgen}. As the auxiliary information for this user
intervention, the \textbf{Polycube Boundary Surface Construction
  Algorithm} will output corners and connectivity of the segmented
surface patches into \textit{.k} file.  Finally, the generated polycube
structure is the combination of non-degenerated cubes and
degenerated cubes splitting the volumetric domain of the geometry.

\subsection{Parametric mapping and subdivision}
\label{sec:octree_subdivision}

After the polycube is constructed, we need to build a bijective
mapping between the input triangle mesh and the boundary surface of
the polycube structure. In our software, we implement the same idea as
in~\cite{Liu2015}: using a non-degenerated unit cube or a degenerated
unit cube as the parametric domain for the polycube structure. As a
result, we can construct a generalized polycube structure that can better
align with the given geometry and generate a high quality
hex-dominant mesh.

There are three types of elements in the hex-dominant mesh: hex,
prism, and tetrahedral. The hex elements form non-degenerated
cubic regions. Prism and tetrahedral elements form degenerated
cubic regions.  We will use octree subdivision to generate hex
elements for non-degenerated cubic regions, while using subdivision to
generate prism and tetrahedral elements for degenerated cubic
regions. Through the pseudocode in the \textbf{Parametric Mapping
  Algorithm} in~\cite{yu2020hexgen}, we describe how to combine the
segmented surface mesh, the polycube structure, and the unit cube to
create an all-hex mesh. We use this algorithm to create the hex
elements in non-degenerated cubic regions. Each non-degenerated cube
in the polycube structure represents one volumetric region of the
geometry and has a non-degenerated unit cube as its parametric
domain. Region \textit{A} in Fig.~\ref{fig:Pipeline_Structure}(d, f)
shows an example of a non-degenerated cube and its corresponding
volume domain of the geometry marked in the green circle.
For degenerated cubes, there are two types of interface, a triangular
face and a quadrilateral face. Region \textit{B} in
Fig.~\ref{fig:Pipeline_Structure}(d, f) shows a prism case, it contains
two triangular faces and three quadrilateral faces. For the tetrahedral case
shown in Region \textit{C} in Fig.~\ref{fig:Pipeline_Structure}(d, f), it
contains four triangular faces.

\begin{figure}[htp]
\centering
\begin{tabular}{cc}
\includegraphics[height=0.6\linewidth]{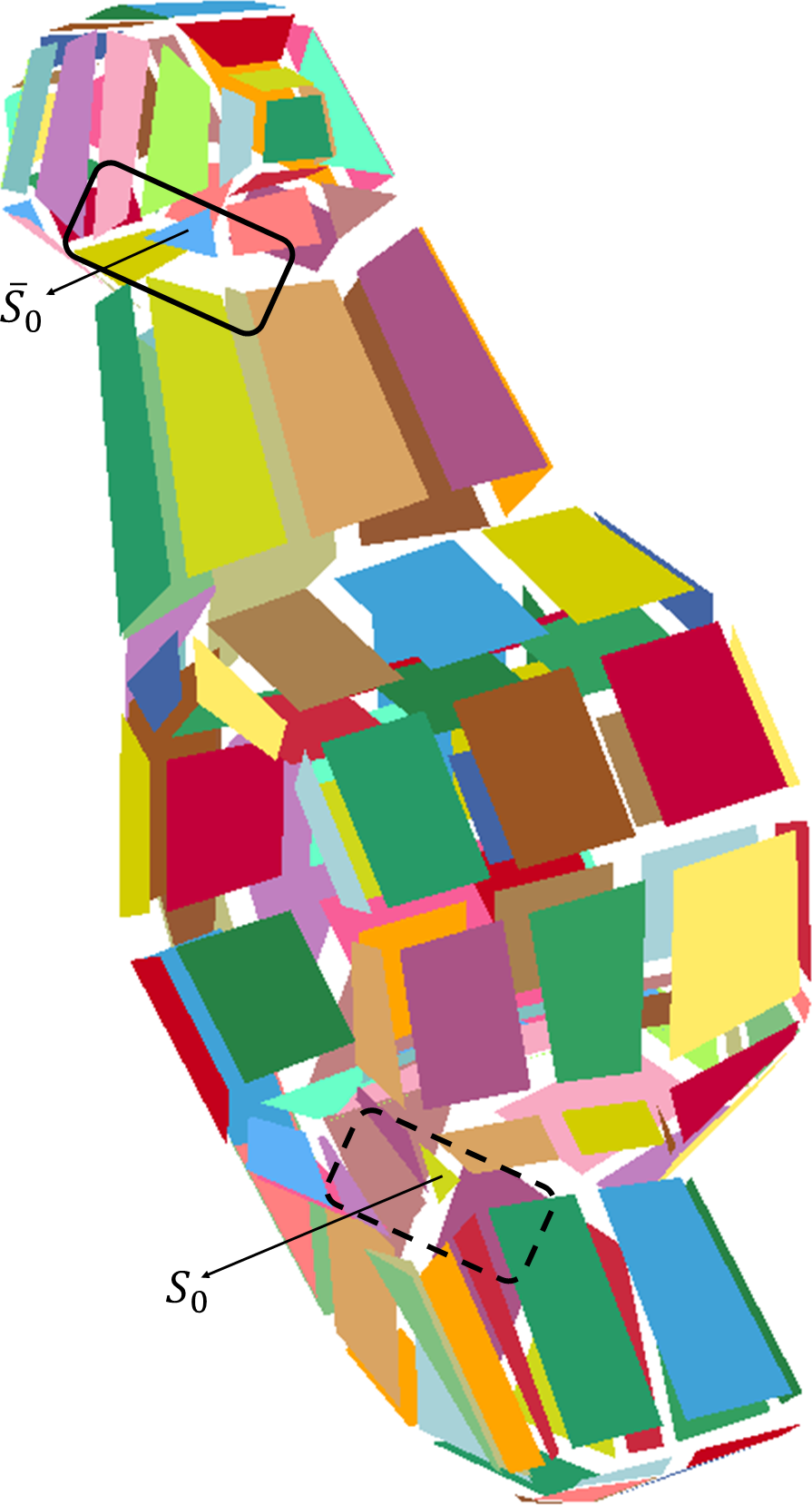}&
\includegraphics[height=0.6\linewidth]{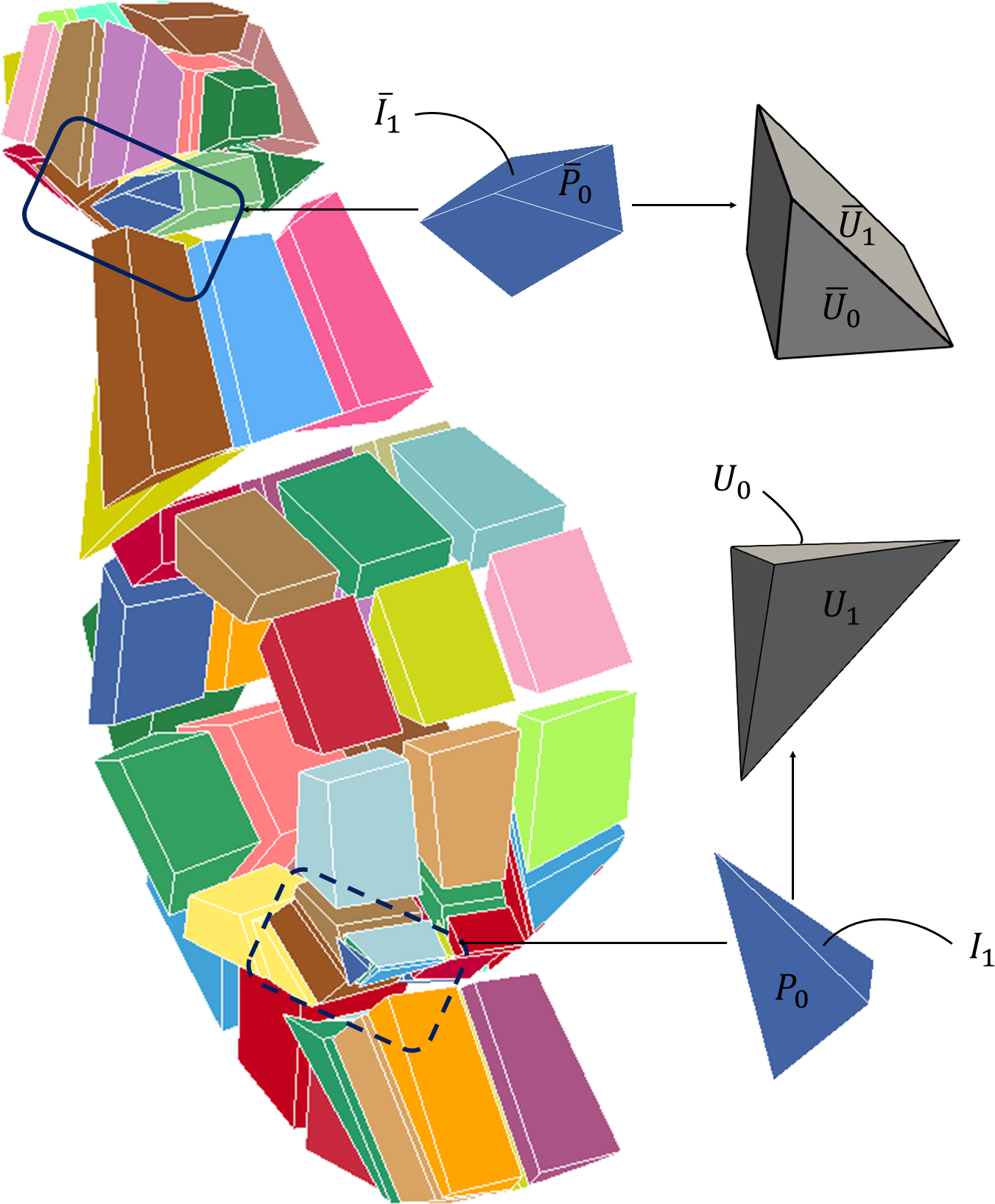}\\
(a) & (b)\\
\end{tabular}
\caption{The polycube construction and the parametric mapping process
  for prism-shape degenerated cubic regions (see the black boxes) and tetrahedral-shape
  degenerated cubic regions (see the dashed black boxes). (a) The boundary surface of the polycube
  generated by \textbf{Polycube Boundary Surface Construction
    Algorithm}; (b) $\bar{S}_0$, $\bar{P}_0$ and $\bar{U}_0$ are used
  for parametric mapping to create boundary vertices of the
  prism-shape degenerated cubic regions. $\bar{I}_1$ and $\bar{U}_1$
  are used for linear interpolation to create interior vertices of the
  prism-shape degenerated cubic regions.  $S_0$, $P_0$ and $U_0$ are
  used for parametric mapping to create boundary vertices of the
  tetrahedral-shape degenerated cubic regions. $I_1$ and $U_1$ are
  used for linear interpolation to create interior vertices of the
  tetrahedral-shape degenerated cubic regions.  }
    \label{fig:prism_polycube_construction}
\end{figure}

Through the pseudocode in the~\nameref{alg:3}, we describe how the
segmented surface mesh, the polycube structure and the prism-shape
degenerated unit cube are combined to generate prism elements. Let
$\{\bar{S}_i\}_{i=1}^N$ be the segmented surface patches coming from
the segmentation result (see
Fig.~\ref{fig:prism_polycube_construction}(a)).  Each segmented
surface patch corresponds to one boundary surface of the polycube
$\bar{P}_i$ $(1\leq i \leq N)$ (see
Fig.~\ref{fig:prism_polycube_construction}(b)), where $N$ is the
number of the boundary surfaces. There are also interior surfaces,
denoted by $\bar{I}_j$ $(1\leq j \leq M)$, where $M$ is the number of the
interior surfaces. The union of $\{\bar{P}_i\}_{i=1}^N$ and
$\{\bar{I}_j\}_{j=1}^M$ is the set of surfaces of the polycube
structure. For the parametric domain, let $\{\bar{U}_k\}_{k=1}^5$
denote the five surface patches of the prism-shape degenerated unit
cube (see Fig.~\ref{fig:prism_polycube_construction}(b)).

Each prism-shape degenerated cube in the polycube structure represents
one volumetric region of the geometry and has a prism-shape
degenerated unit cube as its parametric
domain. Fig.~\ref{fig:prism_polycube_construction}(b) shows an example
of prism-shape degenerated cube and its corresponding volume domain of
the geometry marked in the black boxes. Therefore, for each prism-shape
degenerated cube in the polycube structure, we can find its boundary
surface $\bar{P}_i$ and map the segmented surface patch $\bar{S}_i$ to
its corresponding parametric surface $\bar{U}_k$ of the prism-shape
degenerated unit cube. To map $\bar{S}_i$ to $\bar{U}_k$, we first map
its corresponding boundary edges of $\bar{S}_i$ to the boundary edges
of $\bar{U}_k$. Then we get the parameterization of $\bar{S}_i$ by
using the cotangent Laplace operator to compute the harmonic
function~\cite{wenyan2011b,eck1995multiresolution}. Compared to
non-degenerated cubic regions algorithm, we introduce three parametric
variables in mapping since one face is not axis-aligned. Note that for
an interior surface $\bar{I}_j$ of the polycube structure, we skip the
parametric mapping step.

\begin{figure}[htp]
  \centering
  \begin{tabular}{ccc}
    \includegraphics[height=0.3\linewidth]{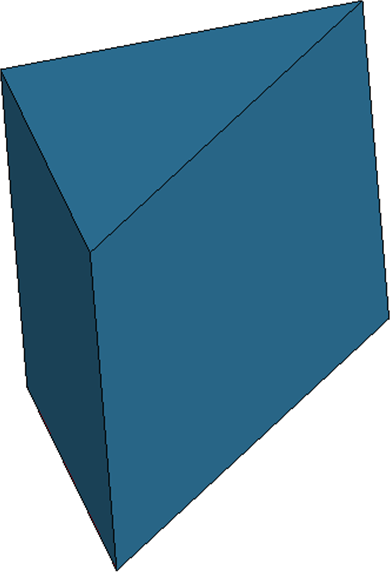}&
    \includegraphics[height=0.3\linewidth]{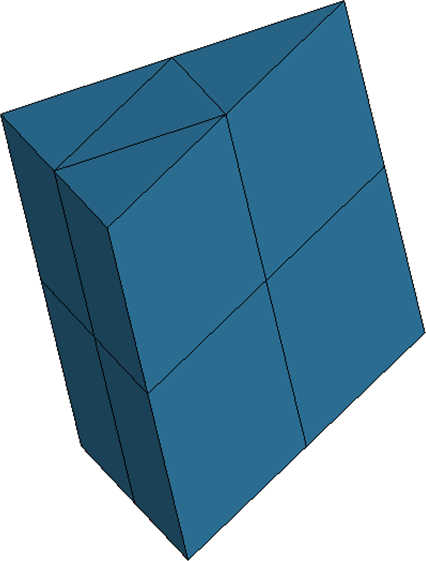}&
    \includegraphics[height=0.3\linewidth]{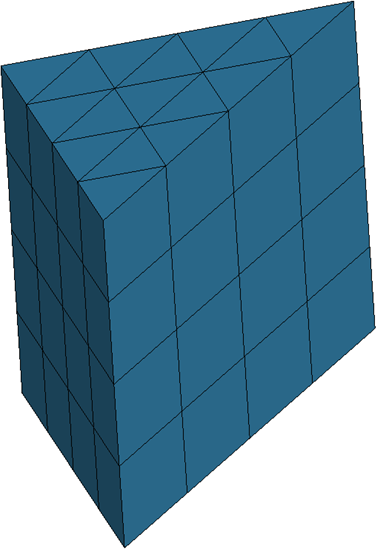}\\
    \includegraphics[height=0.23\linewidth]{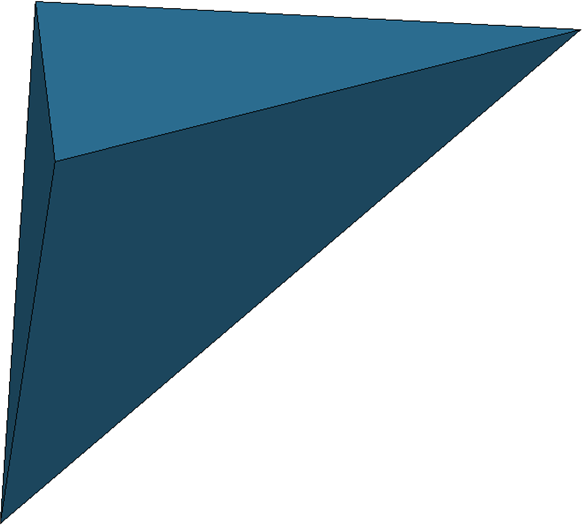}&
    \includegraphics[height=0.23\linewidth]{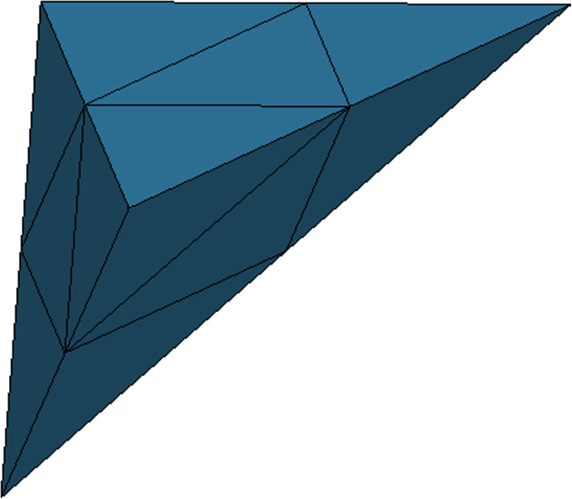}&
    \includegraphics[height=0.23\linewidth]{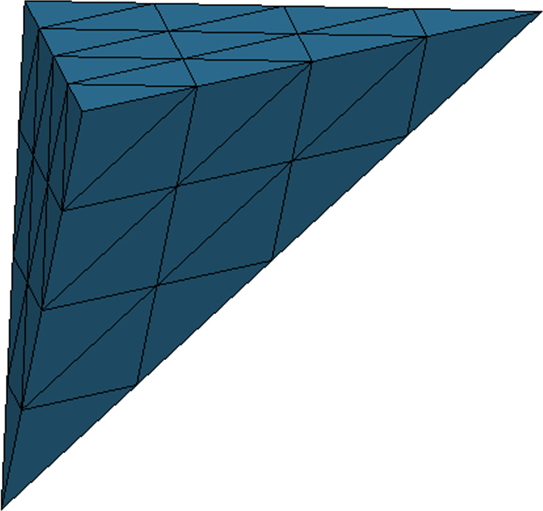}\\
    (a) & (b) & (c)\\
  \end{tabular}
  \caption{The subdivision of prism-shape degenerated unit cube (top row) and
    tetrahedral-shape degenerated unit
    cube (bottom row). (a) Subdivision level 0; (b) subdivision level
    1; and (c) subdivision level 2.}\label{fig:subdivision_degenerated}
\end{figure}

The prism elements can then be obtained from the above surface
parameterization combined with subdivision. We generate the prism
elements for each prism-shape region in the following process. To
obtain vertex coordinates on the segmented patch $\bar{S}_i$, we first
subdivide the prism-shape degenerated unit cube (see
Fig.~\ref{fig:subdivision_degenerated}(a)) recursively in order to get
their parametric coordinates. The vertex coordinates of triangular
faces of the prism-shape degenerated cube are obtained by linear
subdivison, while the quadrilateral faces are also obtained by linear
subdivision. The physical coordinates can be obtained by using
parametric mapping, which has a one-to-one correspondence between the
parametric domain $\bar{U}_k$ and the physical domain $\bar{S}_i$. To
obtain vertices on the interior surface of the prism region, we
skip the parametric mapping step and directly use linear interpolation
to calculate the physical coordinates. Finally, vertices inside
the cubic region are calculated by linear interpolation. The entire
prism elements are built by going through all the prism-shape regions.

\begin{algorithm}[H]
  \caption{\textbf{Prism Parametric Mapping Algorithm}}
  \label{alg:3}
  \begin{algorithmic}[1]
    \Require Segmented triangle mesh
    $\{\bar{S}_i\}_{i=1}^N$, polycube structure
    \Ensure Prism elements in prism-shape degenerated cubic regions
    \State Find boundary surfaces $\{\bar{P}_i\}_{i=1}^N$ and
    interior surfaces $\{\bar{I}_j\}_{j=1}^M$ in the polycube structure
    \renewcommand{\algorithmicensure}{\textbf{Surface parameterization step:}}
    \Ensure

    \For{each prism-shape degenerated cube in the polycube structure}
    \State Create a prism-shape degenerated cube region $\{\bar{U}_k\}_{k=1}^5$ as the parametric domain
    \For{each surface in the prism-shape degenerated cube}
    \If {it is a boundary surface $\bar{P}_i$}
    \If {the surface is not axis-aligned}
    \State Get the surface parameterization $f: \bar{S}_i \to \bar{U}_k \subset \mathbb{R}^3$
    \Else
    \State Get the surface parameterization $f: \bar{S}_i \to \bar{U}_k \subset \mathbb{R}^2$
    \EndIf
    \EndIf
    \EndFor
    \EndFor
    \renewcommand{\algorithmicensure}{\textbf{Parametric mapping and subdivision step:}}
    \Ensure
    \For{each prism-shape degenerated cube in the polycube structure}
    \State Subdivide the prism-shape degenerated unit cube recursively to get parametric coordinates $v_{_{para}}$
    \For{each surface in the prism-shape degenerated cube}
    \If {it is a boundary surface $\bar{P}_i$}
    \State Obtain physical coordinates using $f^{-1}(v_{_{para}})$
    \ElsIf {it is an interior surface $\bar{I}_j$}
    \State Obtain physical coordinates using linear interpolation
    \EndIf
    \EndFor
    \State Obtain interior vertices in the prism-shape degenerated cubic region using linear interpolation
    \EndFor
  \end{algorithmic}
\end{algorithm}

We perform the similar procedure for the tetrahedra-shape degenerated
cubes in the polycube structure. Through the pseudocode in
the~\nameref{alg:4}, we describe how the segmented surface mesh, the
polycube structure and the tetrahedral-shape degenerated unit cube are
combined to generate tetrahedral
elements. Fig.~\ref{fig:prism_polycube_construction}(b) shows an
example of tetrahedral-shape degenerated cube and its corresponding
volume domain of the geometry marked in the dashed black boxes. The
difference is that we use $\{U_k\}_{k=1}^4$ to denote those four
surface patches of the tetrahedra-shape degenerated unit cube for the
parametric domain. We also introduce three parametric variables in
mapping when one of the surfaces is not axis aligned. Then, the
tetrahedral elements can be obtained from this surface
parameterization combined with linear subdivision. We generate
tetrahedral elements for each tetrahedral-shape region in the
following process. To obtain vertex coordinates on the segmented patch
$S_i$, we first subdivide the tetrahedral-shape degenerated unit cube
(see Fig.~\ref{fig:subdivision_degenerated}(bottow row)) recursively
in order to get their parametric coordinates by applying linear
subdivison. The physical coordinates can be obtained by using the
parametric mapping between the parametric domain $U_k$ and the
physical domain $S_i$.  $I_1$ and $U_1$ are combined for linear
interpolation to obtain vertices on the interior surface of the
tetrahedra-shape degenerated cubic region. Finally, vertices
inside the tetrahedra-shape degenerated cube region are calculated by
linear interpolation. The entire tetrahedral elements are built by
going through all the tetrahedral regions.

\begin{algorithm}[H]
  \caption{\textbf{Tetrahedral Parametric Mapping Algorithm}}
  \label{alg:4}
        \begin{algorithmic}[1]
            \Require Segmented triangle mesh
            $\{S_i\}_{i=1}^N$, polycube structure
            \Ensure Tetrahedral elements in tetrahedral-shape degenerated cubic regions
            \State Find boundary surfaces $\{P_i\}_{i=1}^N$ and
            interior surfaces $\{I_j\}_{j=1}^M$ in the polycube structure
            \renewcommand{\algorithmicensure}{\textbf{Surface parameterization step:}}
            \Ensure

            \For{each tetrahedral-shape degenerated cube in the polycube structure}
            \State Create a tetrahedral-shape degenerated cube region $\{U_k\}_{k=1}^4$ as the parametric domain
            \For{each surface in the tetrahedral-shape degenerated cube}
               \If {it is a boundary surface $P_i$}
               \If {the surface is not axis-aligned}
               \State Get the surface parameterization $f: S_i \to U_k \subset \mathbb{R}^3$
               \Else
               \State Get the surface parameterization $f: S_i \to U_k \subset \mathbb{R}^2$
               \EndIf
               \EndIf
               \EndFor
               \EndFor
               \renewcommand{\algorithmicensure}{\textbf{Parametric mapping and subdivision step:}}
               \Ensure
               \For{each tetrahedral-shape degenerated cube in the polycube structure}
               \State Subdivide the tetrahedral-shape degenerated unit cube recursively to get parametric coordinates $v_{_{para}}$
               \For{each surface in the tetrahedral-shape degenerated cube}
               \If {it is a boundary surface $P_i$}
               \State Obtain physical coordinates using $f^{-1}(v_{_{para}})$
               \ElsIf {it is an interior surface $I_j$}
               \State Obtain physical coordinates using linear interpolation
               \EndIf
               \EndFor
                \State Obtain interior vertices in the tetrahedral-shape degenerated cubic region using linear interpolation
                \EndFor
        \end{algorithmic}
\end{algorithm}

\begin{figure}[htp]
\centering
\begin{tabular}{cc}
\includegraphics[width=0.46\linewidth]{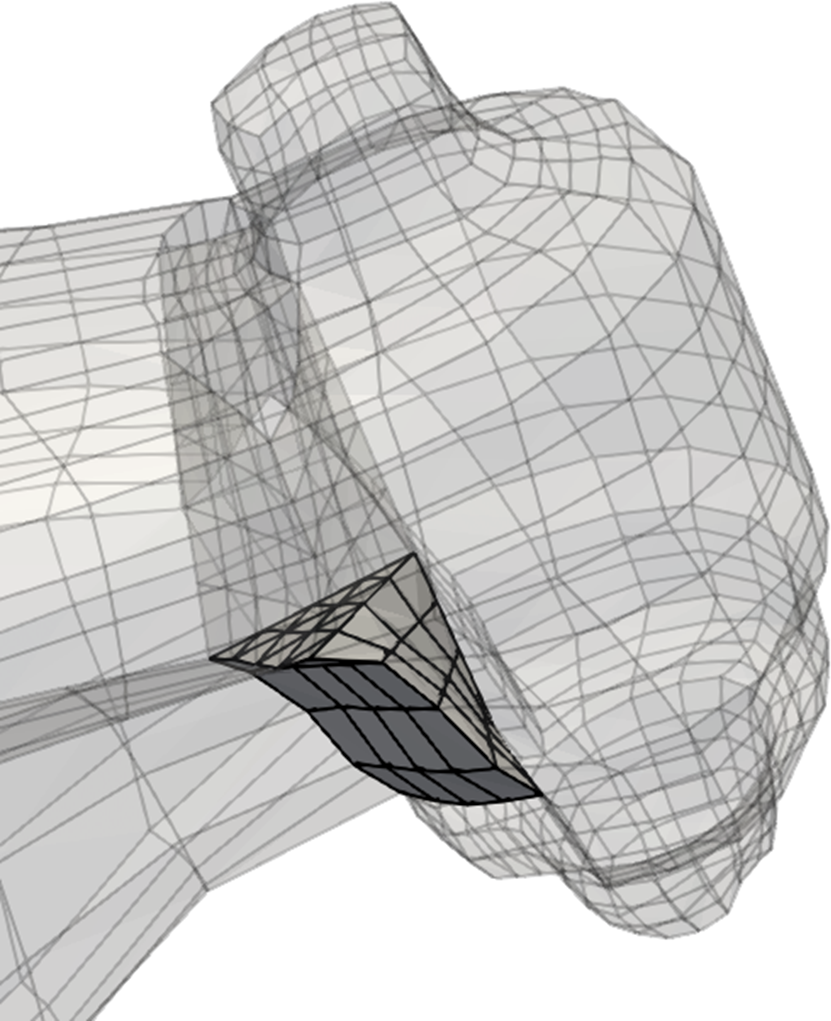}&
\includegraphics[width=0.46\linewidth]{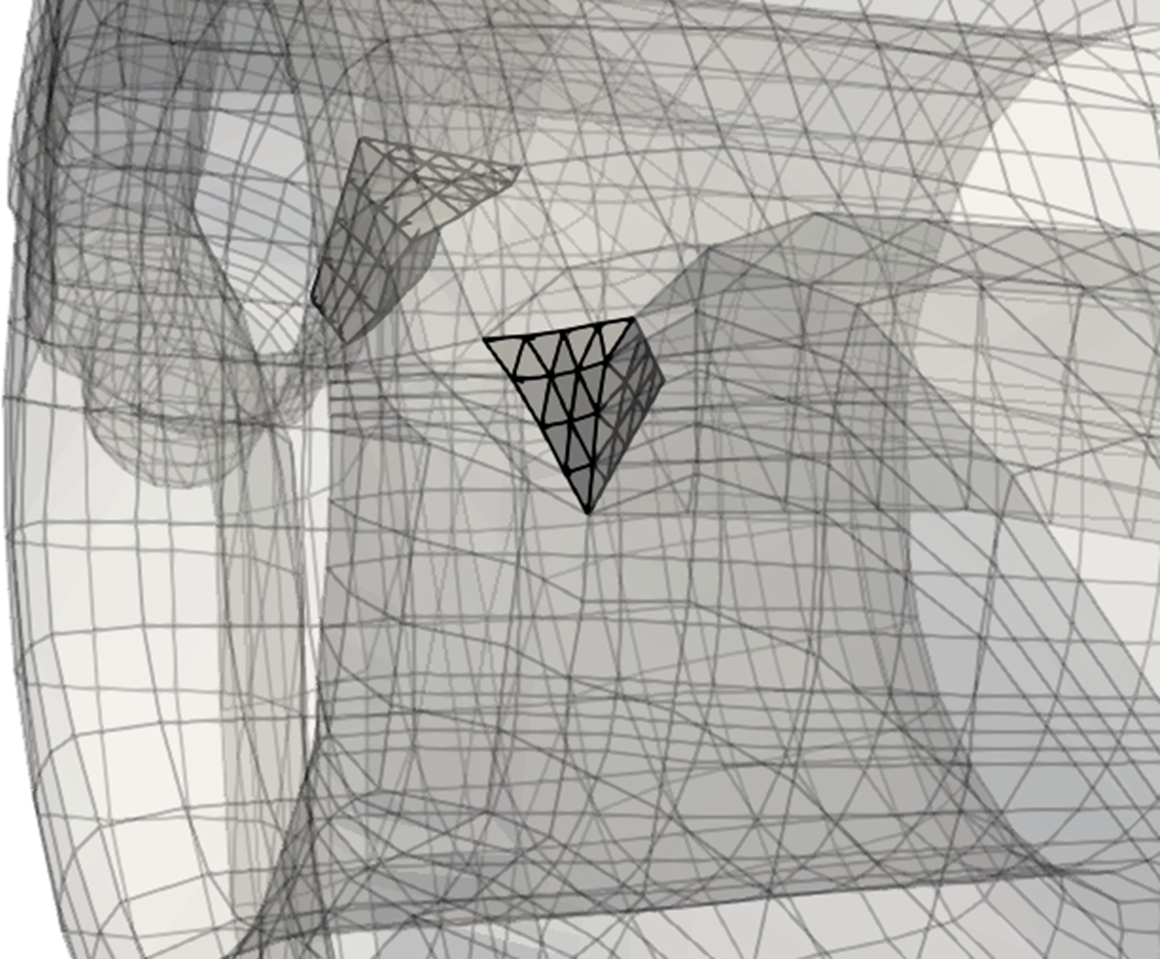}\\
(a) & (b)\\
\end{tabular}
\caption{Prism elements and tetrahedral elements of the rockerarm
  model (some elements are removed to show the interior of
  Fig.~\ref{fig:Pipeline_Structure}(h). (a) Prism elements in
  a prism-shape region; and (b) tetrahedral elements in a
  tetrahedral-shape region.}
    \label{fig:rockerarm_prism}
\end{figure}

Based on the~\nameref{alg:3}, we implemented and organized the code into a
CLI program (PrismGen.exe) that can generate prism elements by
combining parametric mapping with subdivision. Here, we run the
following command to generate the prism elements for the rockerarm
model:
\begin{lstlisting}[style=mystyle]
PrismGen.exe -i rockerarm_indexpatch_read.k -p
rockerarm_polycube_structure.k -o rockerarm_prism.vtk -s 2
\end{lstlisting}
There are four options used in the command:
\begin{itemize}
    \item \textbf{-i}: Surface triangle mesh of the input geometry
      with segmentation information (rockerarm\_indexpatch\_read.k);
    \item \textbf{-o}: Prism mesh (rockerarm\_prism.vtk);
    \item \textbf{-p}: Polycube structure
      (rockerarm\_polycube\_structure.k); and
    \item \textbf{-s}: Subdivision level.
\end{itemize}
We use \textbf{-i} to input the segmentation file generated in
Section~\ref{sec:surface-partition} and use \textbf{-p} to input the
polycube structure created in
Section~\ref{sec:poly_construction}. Option \textbf{-s} is used to set
the level of recursive subdivision. There is no
subdivision if we set \textbf{-s} to be 0. In the rockerarm model, we
set \textbf{-s} to be 2 to create a level-2 prism elements in the
final mesh. The output prism elements are stored in the \textit{vtk} format (see
Fig.~\ref{fig:rockerarm_prism}(a)) and they can be visualized in
Paraview~\cite{ahrens2005paraview}.

Based on~\nameref{alg:4}, we implemented and organized the code into a
CLI program (TetGen.exe) that can generate tetrahedral elements by
combining parametric mapping with linear subdivision. Here, we
run the following command to generate tetrahedral elements for the
rockerarm model:
\begin{lstlisting}[style=mystyle]
TetGen.exe -i rockerarm_indexPatch_read.k -p
rockerarm_polycube_structure.k -o rockerarm_tet.vtk -s 2
\end{lstlisting}
There are five options used in the command:
\begin{itemize}
    \item \textbf{-i}: Surface triangle mesh of the input geometry
      with segmentation information (rockerarm\_indexPatch\_read.k);
    \item \textbf{-o}: Tet mesh (rockerarm\_tet.vtk);
    \item \textbf{-p}: Polycube structure
      (rockerarm\_polycube\_structure.k); and
    \item \textbf{-s}: Subdivision level.
\end{itemize}
We use \textbf{-i} to input the segmentation file generated in
Section~\ref{sec:surface-partition} and use \textbf{-p} to input the
polycube structure created in
Section~\ref{sec:poly_construction}. Option \textbf{-s} is used to set
the level of recursive subdivision. There is no
subdivision if we set \textbf{-s} to be 0. In the rockerarm model, we
set \textbf{-s} to be 2 to create a level-2 tetrahedral mesh. The
output tetrahedral elements are stored in the \textit{vtk} format (see
Fig.~\ref{fig:rockerarm_prism}(b)) and they can be visualized in
Paraview.

\subsection{Quality improvement}

We integrate three quality improvement techniques in the software
package, namely pillowing, smoothing and optimization. Users can
improve mesh quality through the command line options. We first use
pillowing to insert one layer of elements around the
boundary~\cite{wenyan2011b} of the hex elements. By using the
pillowing technique, we ensure that each hex element has at most one
face on the boundary, which can help improve the mesh quality around
the boundary. After pillowing, smoothing and
optimization~\cite{wenyan2011b} are used to further improve the
quality of hex elements. For smoothing, different relocation methods
are applied to three types of vertices: vertices on sharp edges of the
boundary, vertices on the boundary surface, and interior vertices. For
each sharp-edge vertex, we first detect its two neighboring vertices
on the curve, and then calculate their middle point. For each vertex
on the boundary surface, we calculate the area center of its
neighboring boundary quadrilaterals (quads).  For each interior
vertex, we calculate the weighted volume center of its neighboring hex
elements as the new position. We relocate the vertex iteratively. Each
time the vertex moves only a small step towards the new position and
this movement is done only if the new location results in an improved
local Jacobian.
If there are still poor quality hex elements after smoothing, we run
the optimization whose objective function is the Jacobian. Each vertex
is then moved toward an optimal position that maximizes the worst
Jacobian. We presented the \textbf{Quality Improvement Algorithm}
in~\cite{yu2020hexgen} for quality improvement.

\section{HexDom Software and Applications }
The algorithms discussed in Sections~\ref{sec:hexg-polyc-based} were
implemented in C++. The Eigen library~\cite{eigenweb} is used for
matrix and vector operations. We used a compiler-independent building
system (CMake) and a version-control system (Git) to support software
development. We have compiled the source code into the following
software package:
\begin{itemize}
    \item \textbf{HexDom software package}:
    \begin{itemize}
    \item \textbf{Segmentation module (Segmentation.exe);}
    \item \textbf{Polycube construction module (Polycube.exe);}
    \item \textbf{Hex-dominant mesh generation module
        (HexGen.exe, PrismGen.exe, TetGen); and}
    \item \textbf{Quality improvement module (Quality.exe).}
\end{itemize}
\end{itemize}
The software is open-source and can be found in the following Github
link (https://github.com/CMU-CBML/HexDom).

\begin{figure}[htp]
\centering
\begin{tabular}{ccc}
   \includegraphics[height=0.20\textwidth]{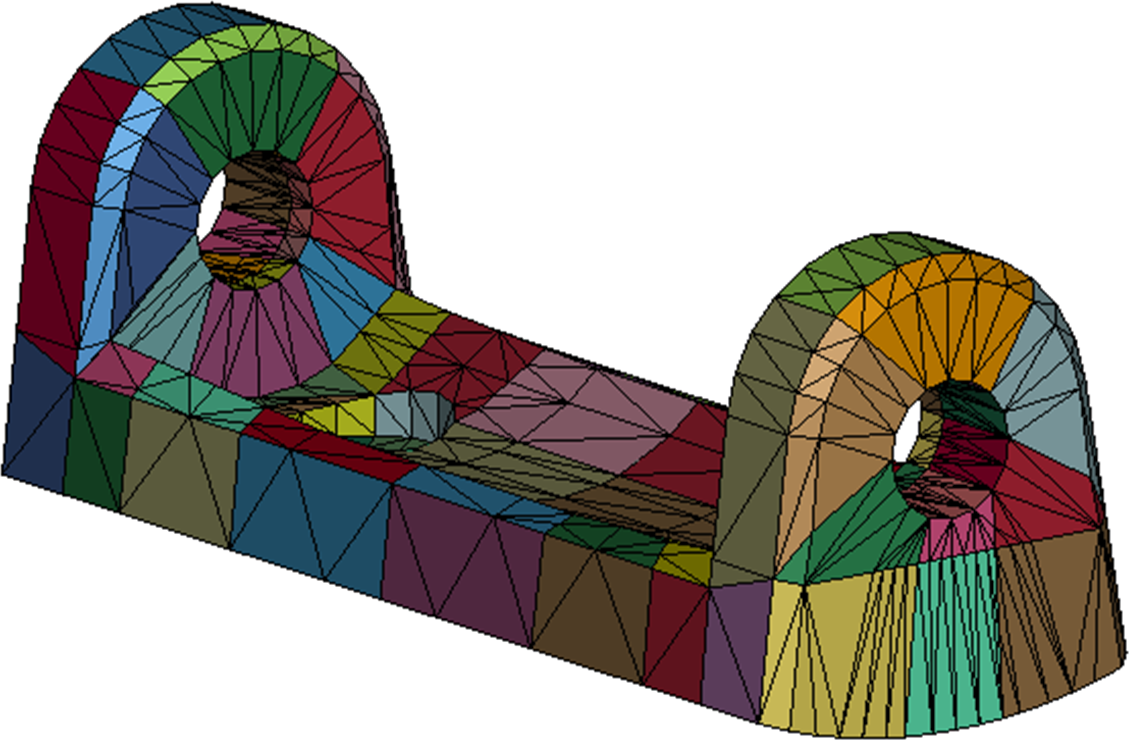}
  &\includegraphics[height=0.20\textwidth]{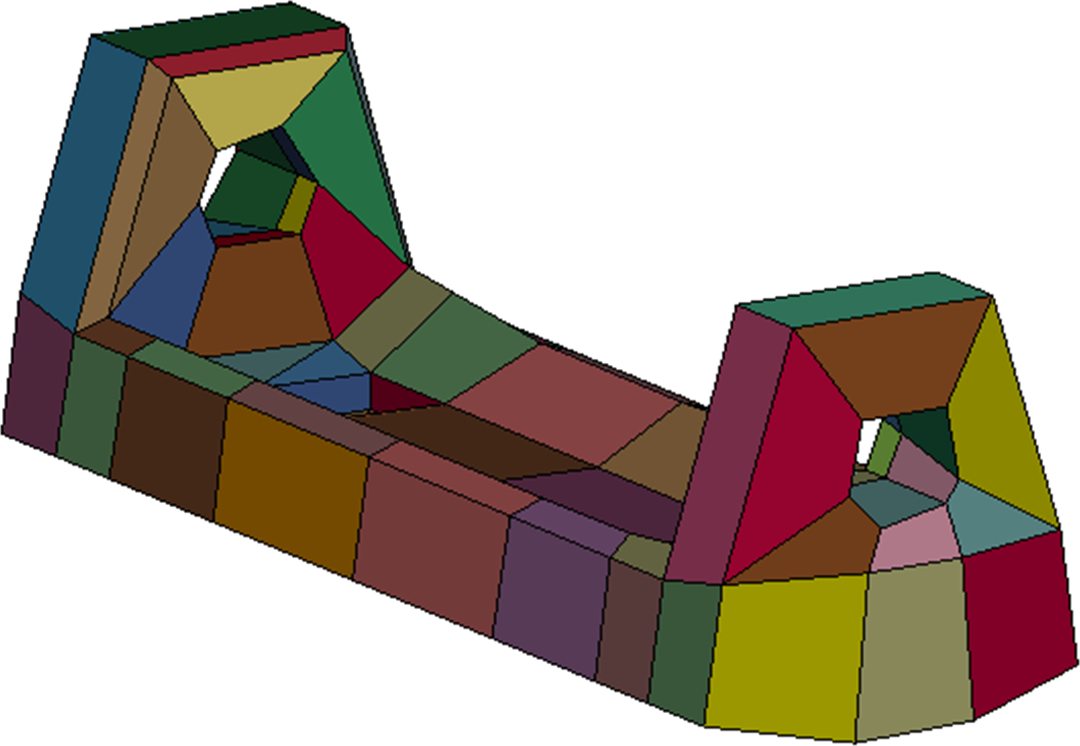}
  &\includegraphics[height=0.20\textwidth]{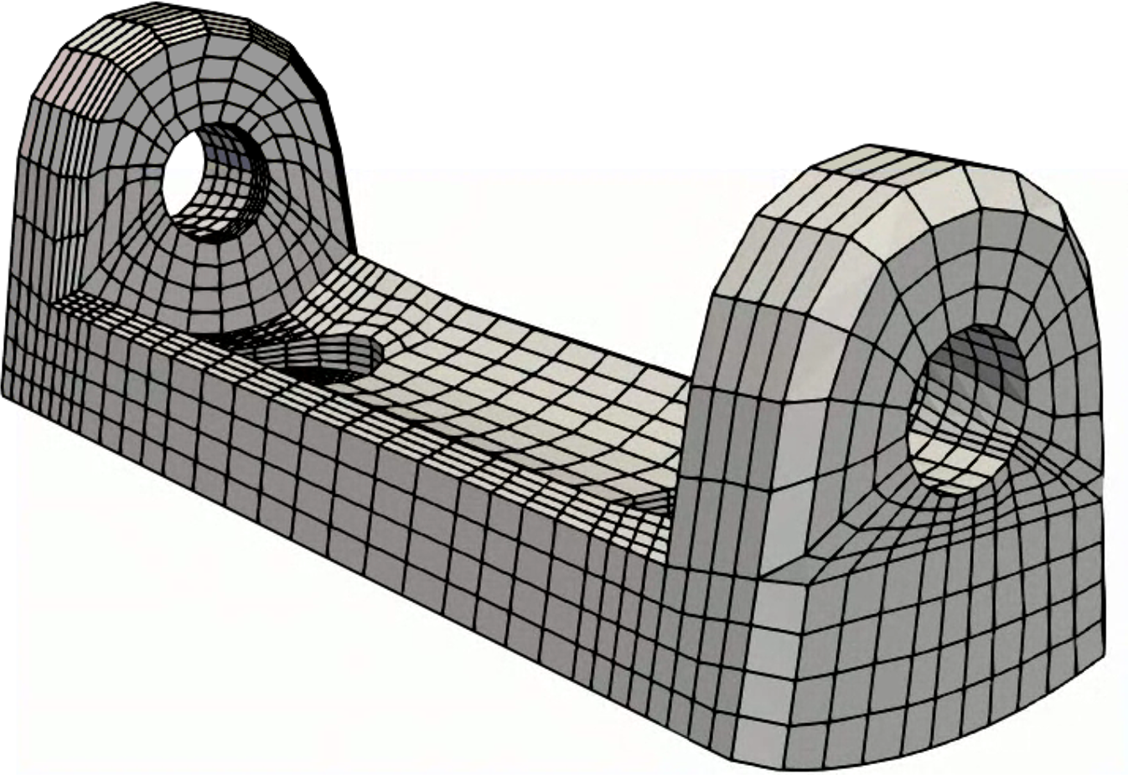}\\
\includegraphics[width=0.33\textwidth]{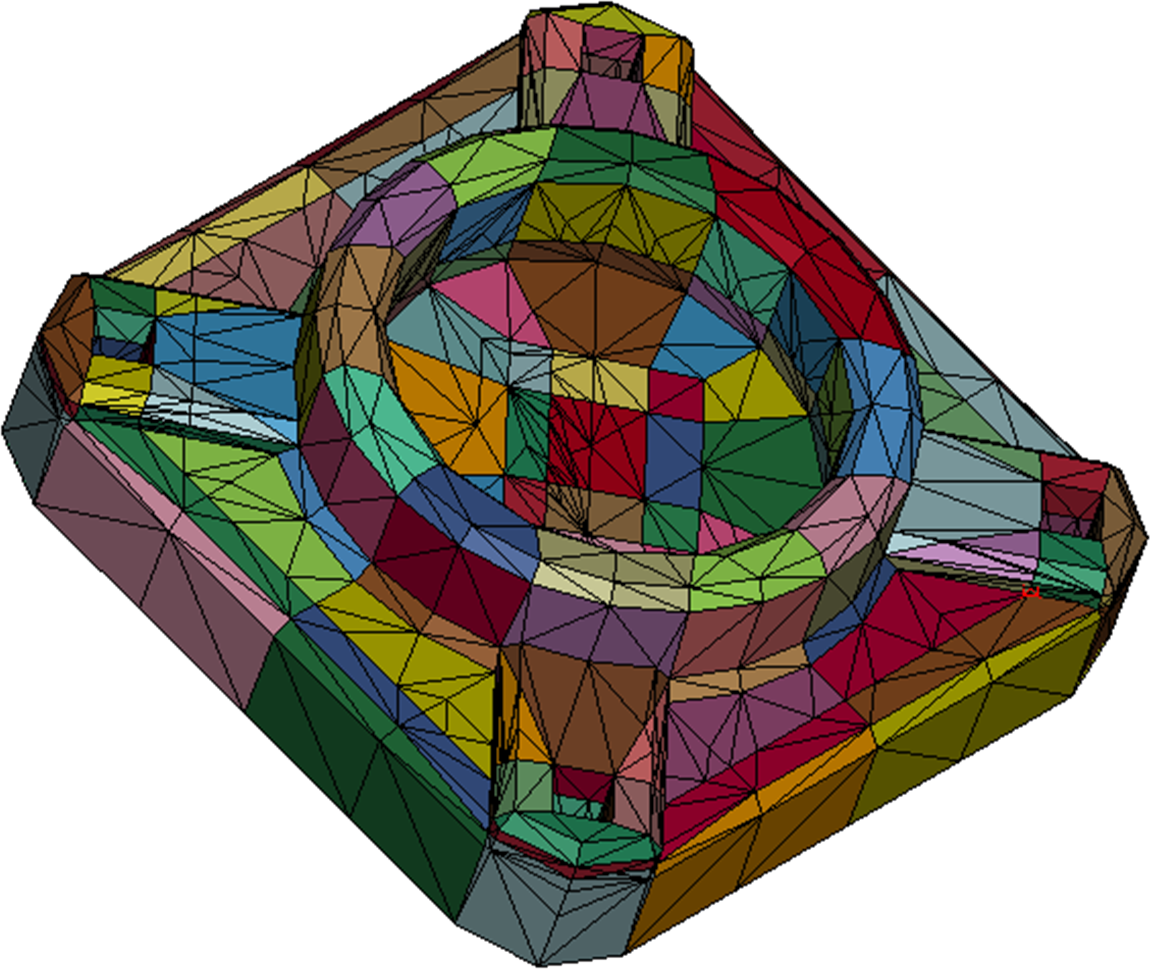}
&\includegraphics[width=0.33\textwidth]{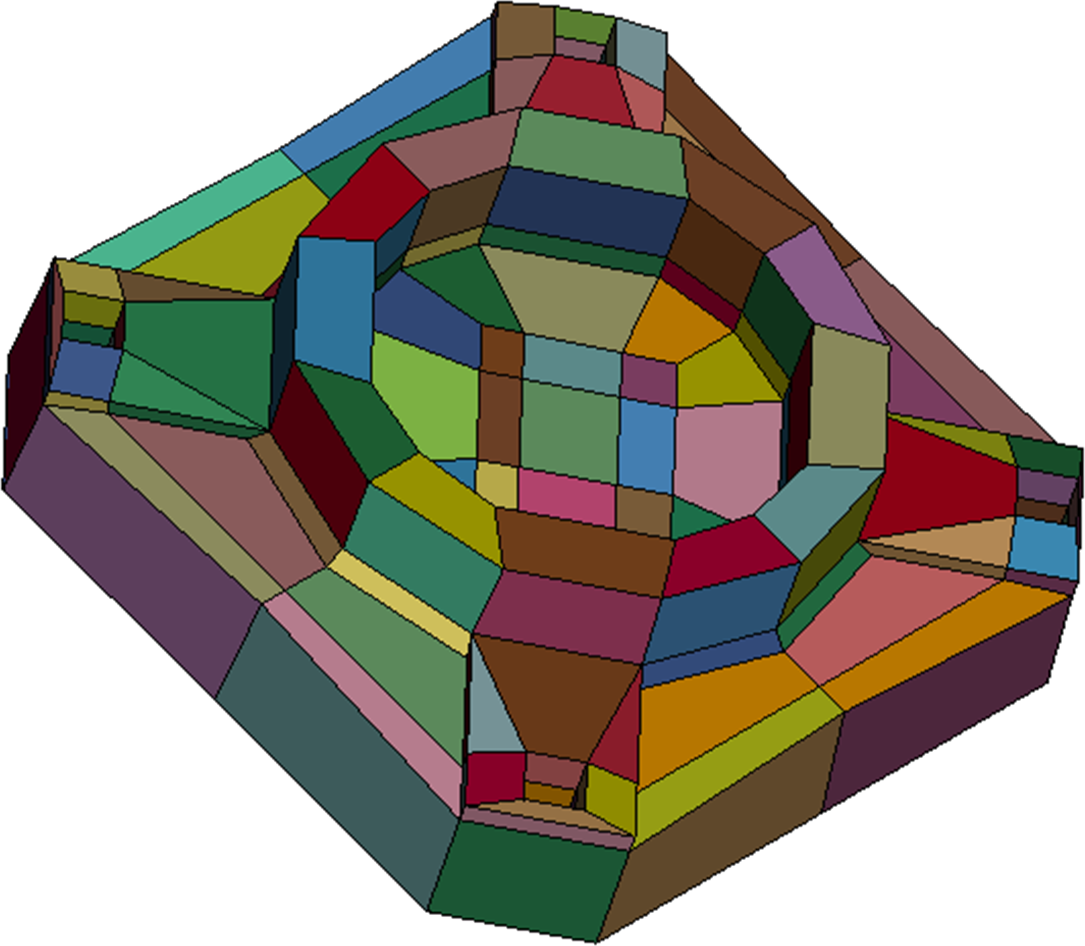}
&\includegraphics[width=0.33\textwidth]{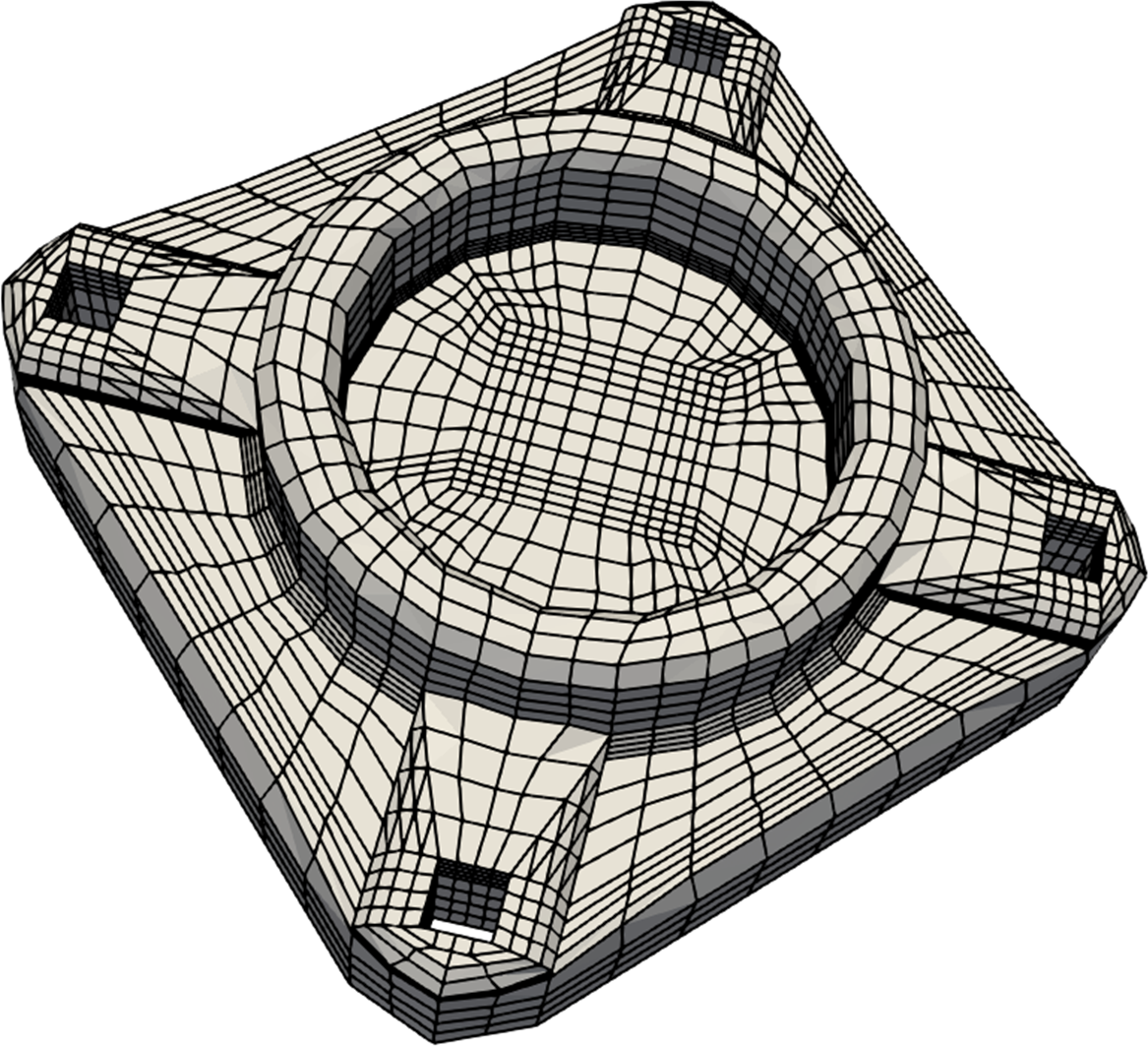}\\
\includegraphics[width=0.33\textwidth]{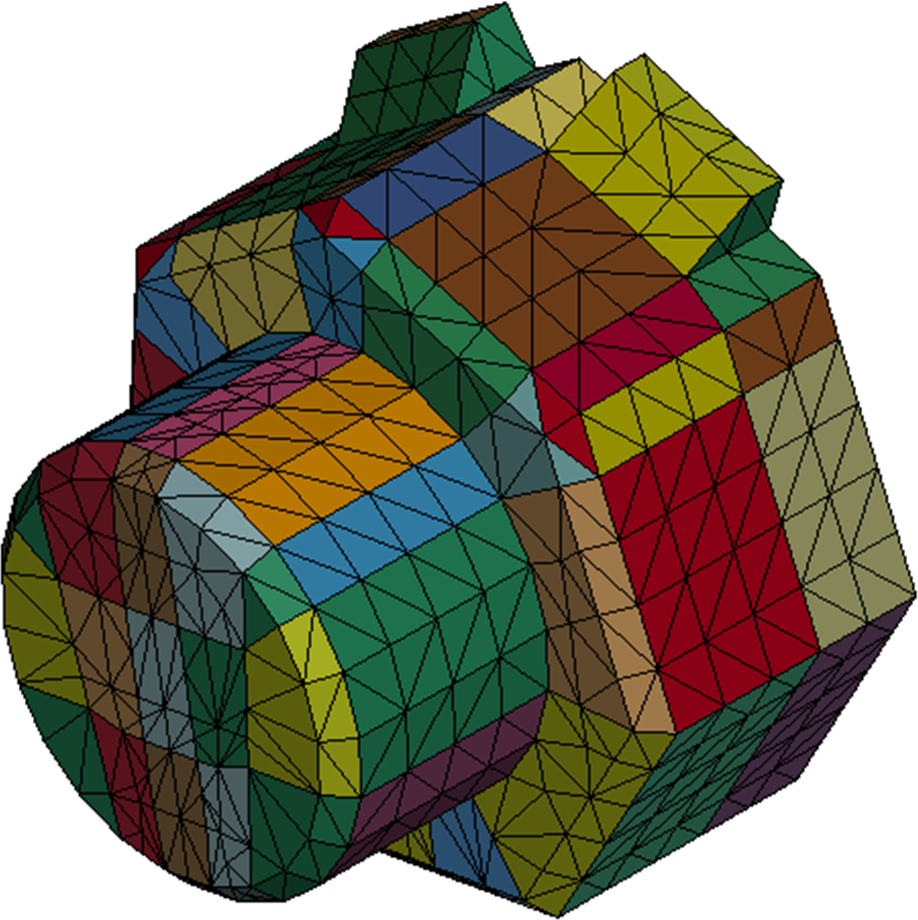}
&\includegraphics[width=0.33\textwidth]{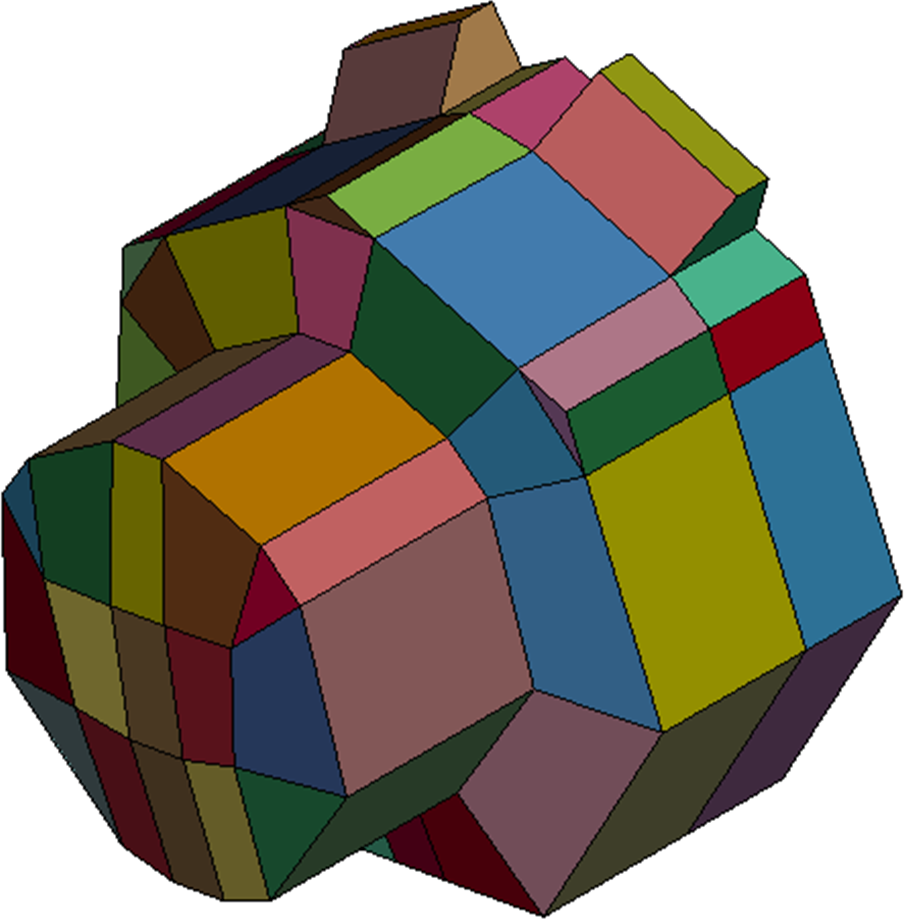}
  &\includegraphics[width=0.33\textwidth]{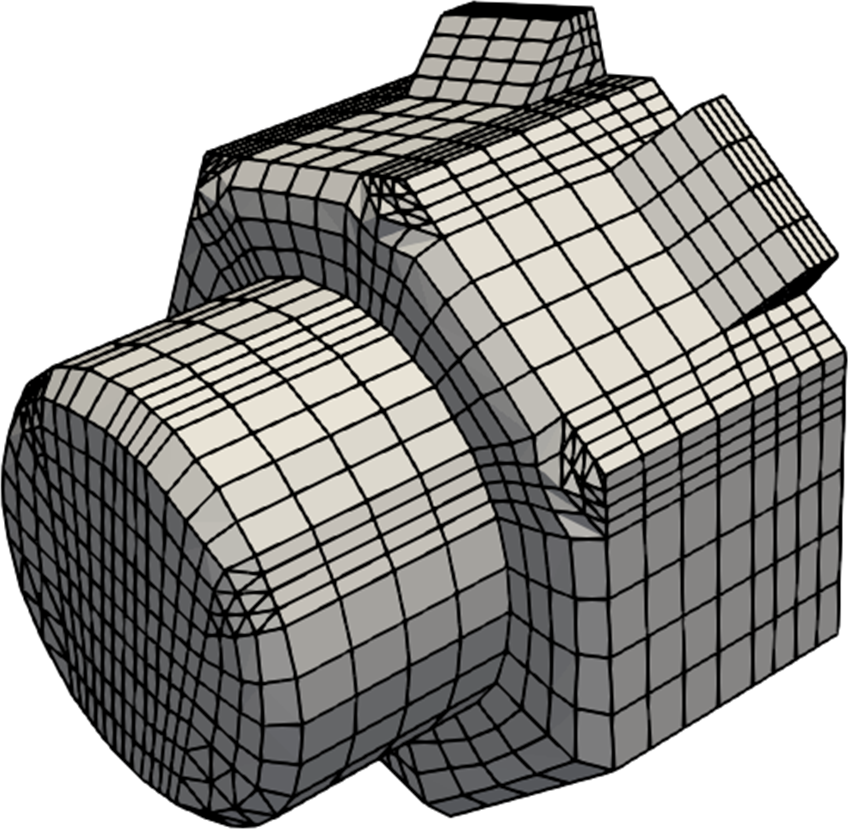}\\
  \includegraphics[height=0.3\linewidth]{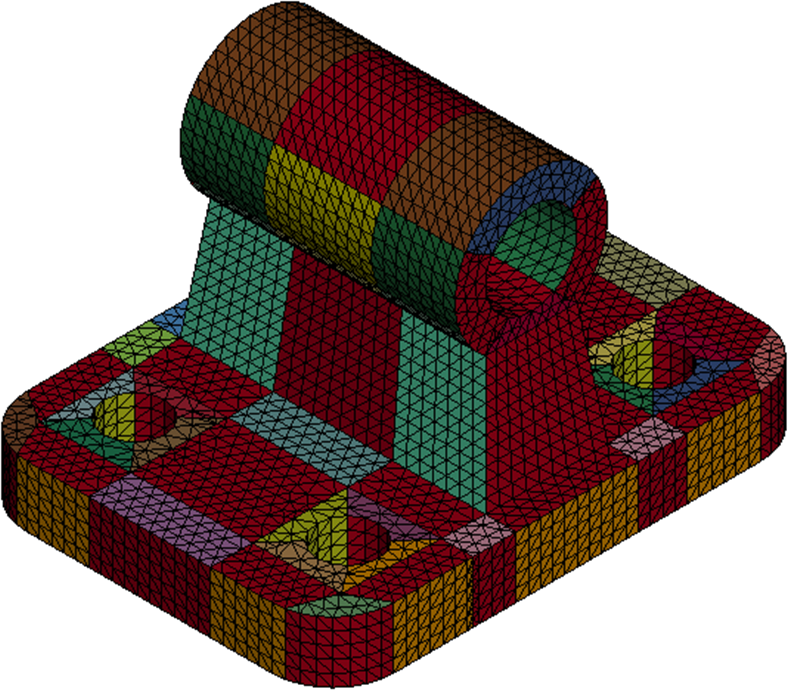}
&\includegraphics[height=0.3\linewidth]{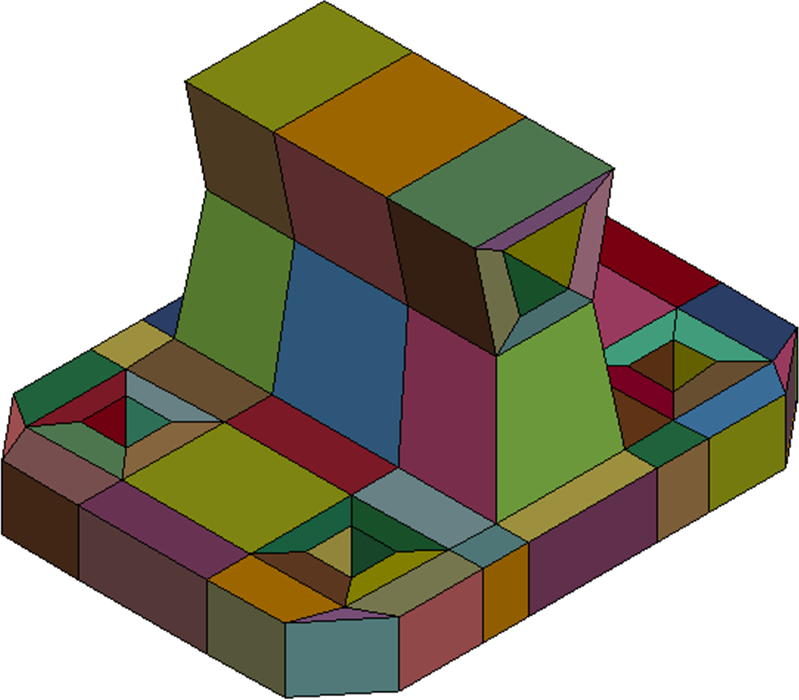}
&\includegraphics[height=0.3\linewidth]{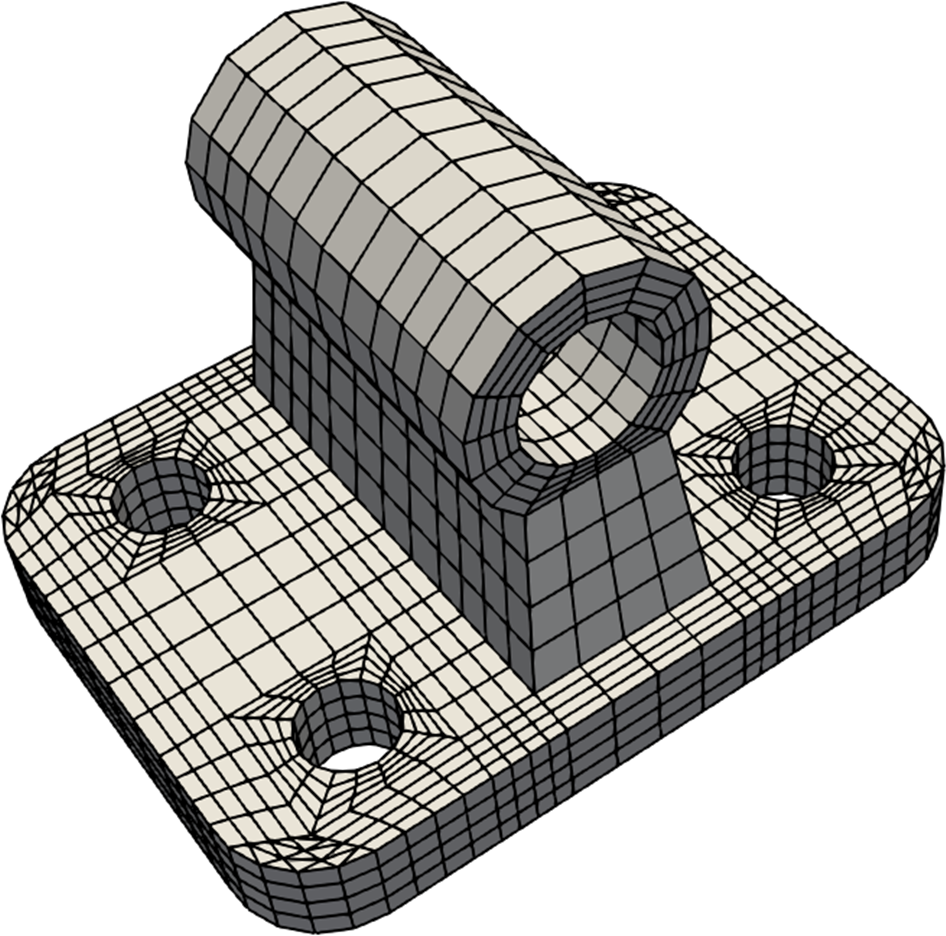}\\
   (a) & (b) & (c)\\
\end{tabular}
\caption{ Results of two types of mount, hepta and a base with four
  holes. (a) Surface triangle meshes and segmentation results; (b)
  polycube structures; and (c) hex-dominant meshes. }
    \label{fig:model1}
  \end{figure}

\begin{figure}[htp]
\centering
\begin{tabular}{ccc}
  \includegraphics[width=0.33\textwidth]{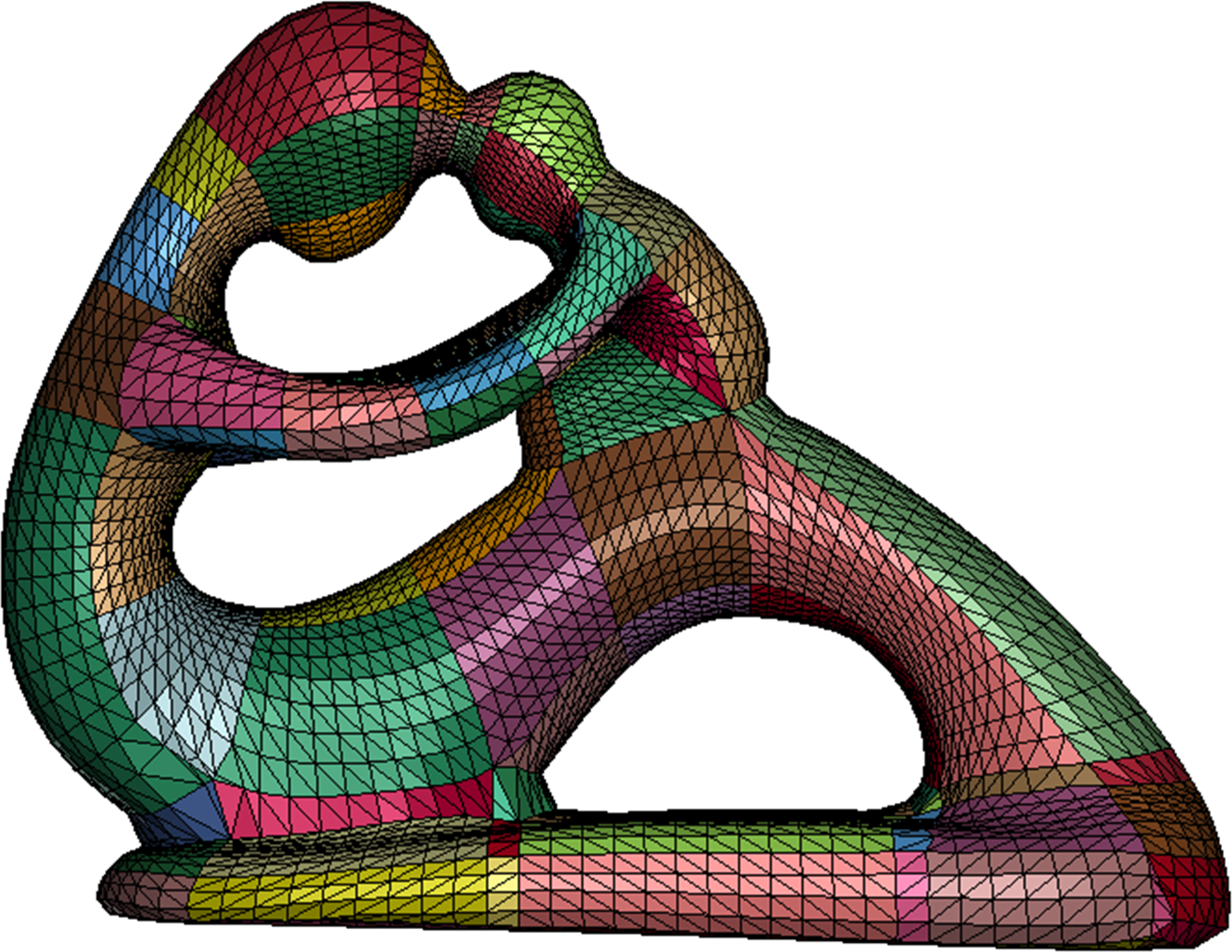}
&\includegraphics[width=0.33\textwidth]{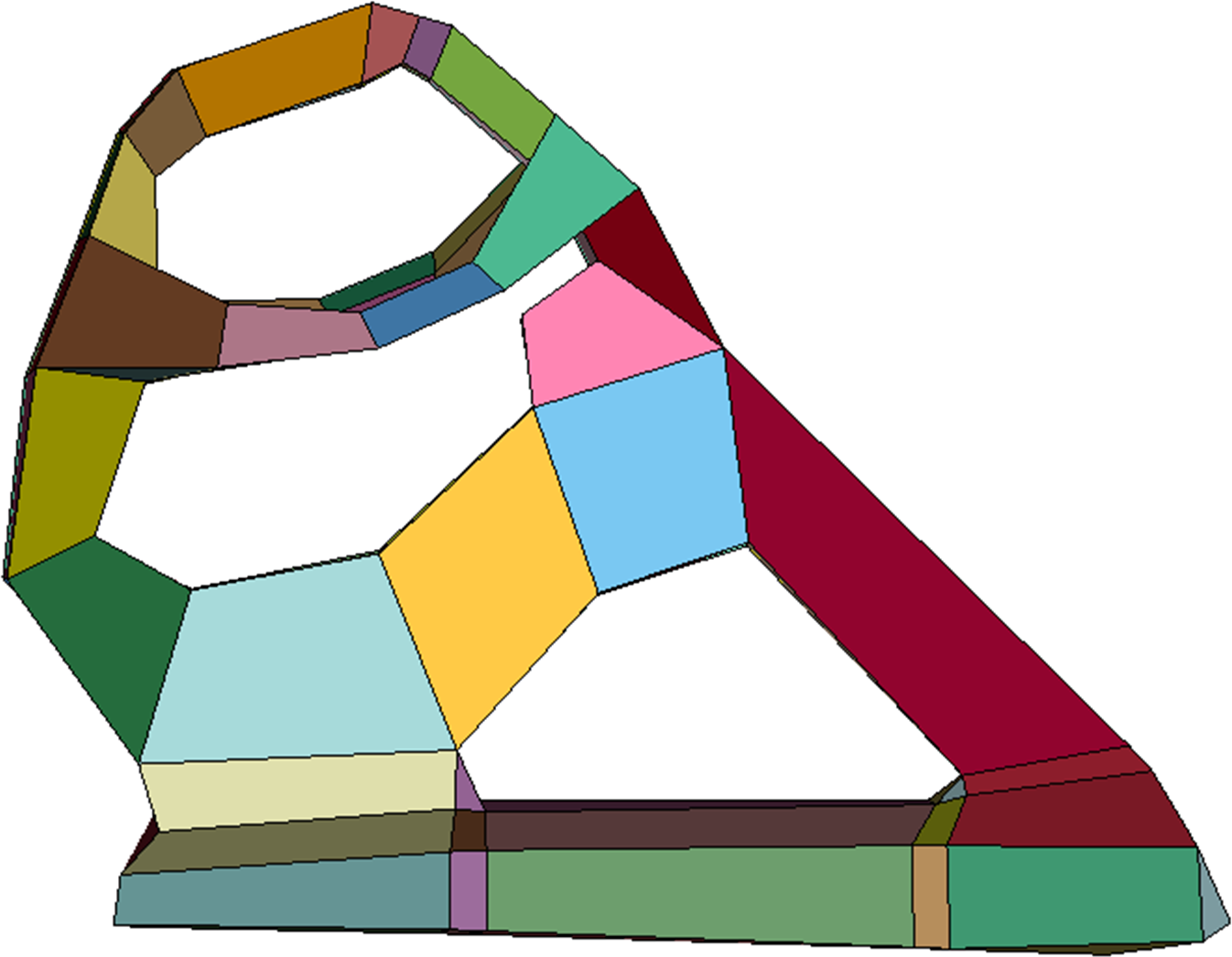}
&\includegraphics[width=0.33\textwidth]{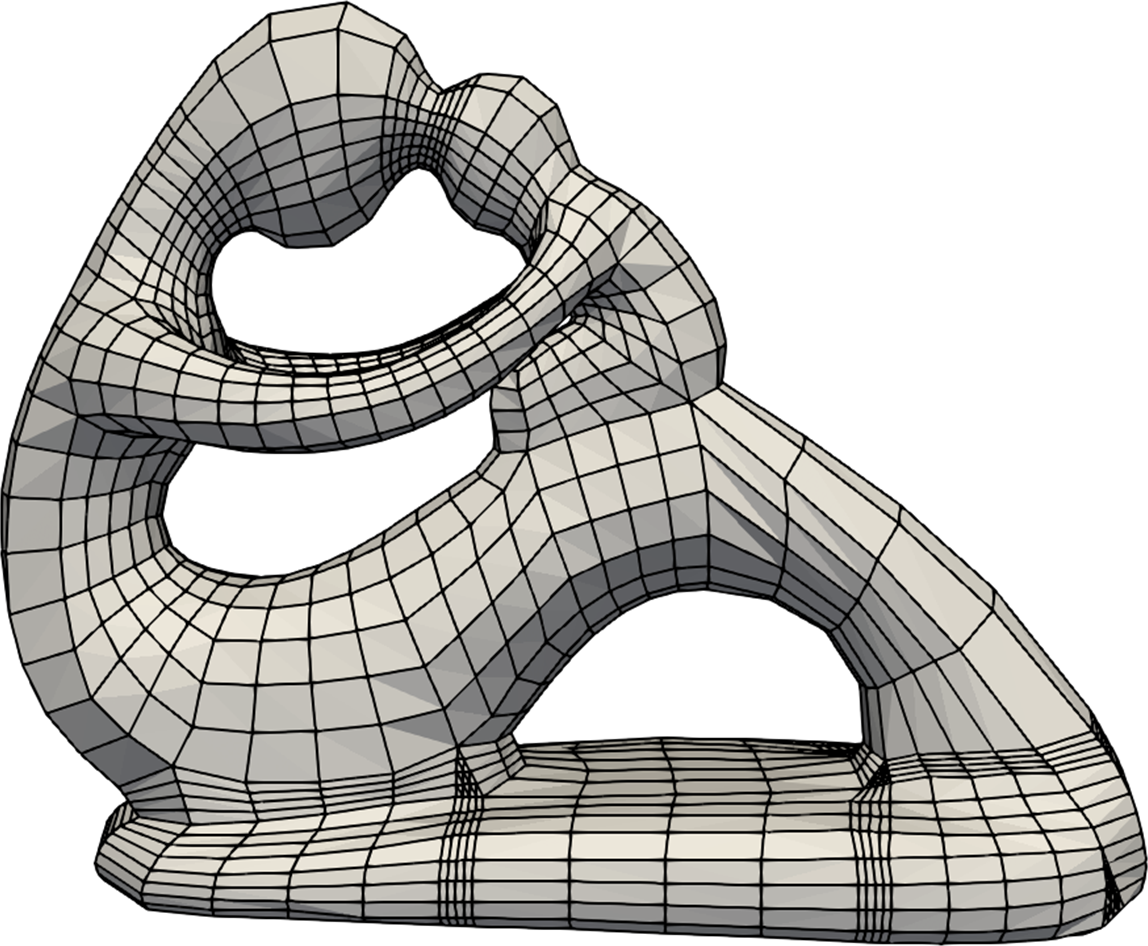}\\
\includegraphics[width=0.33\textwidth]{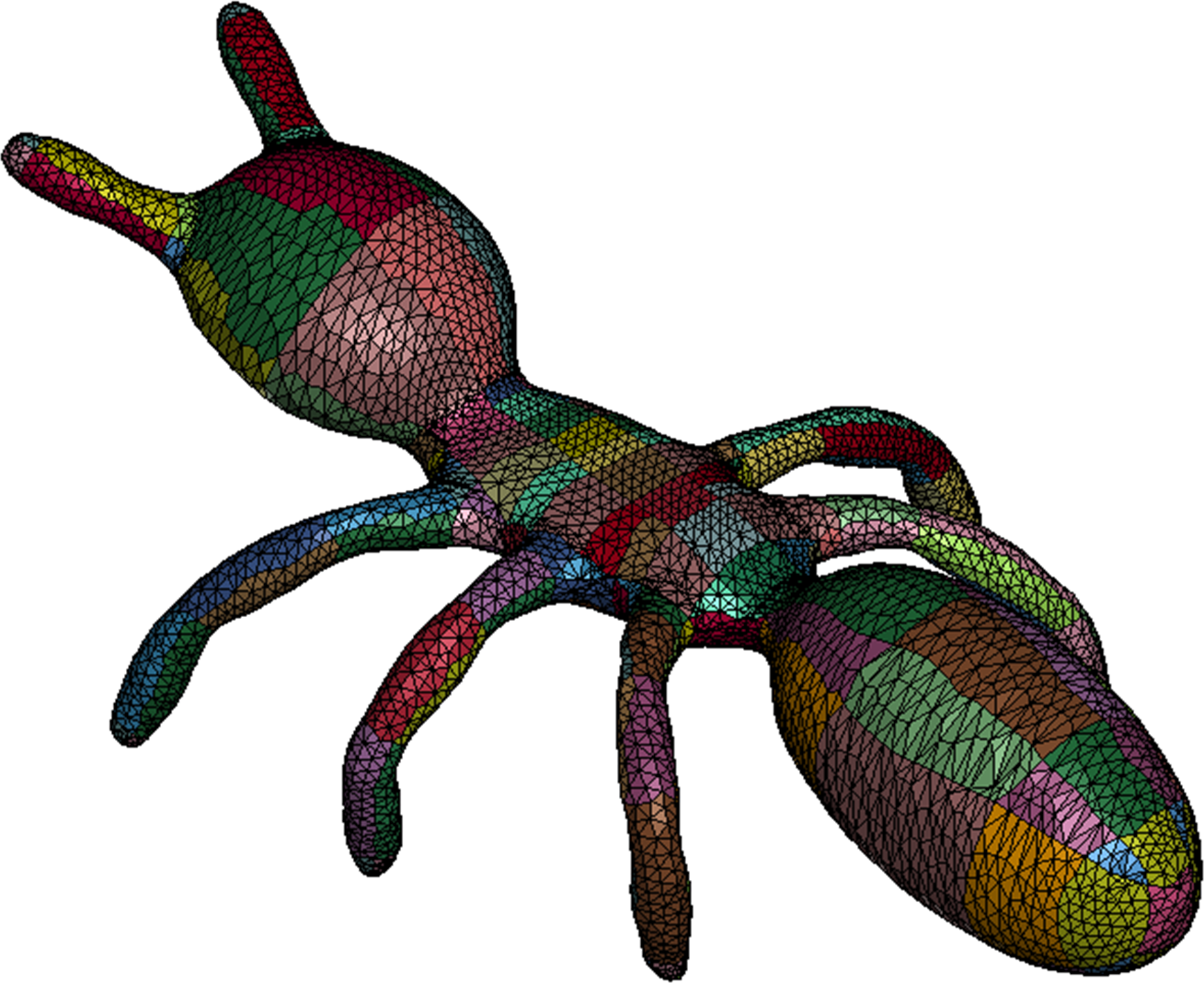}
&\includegraphics[width=0.33\textwidth]{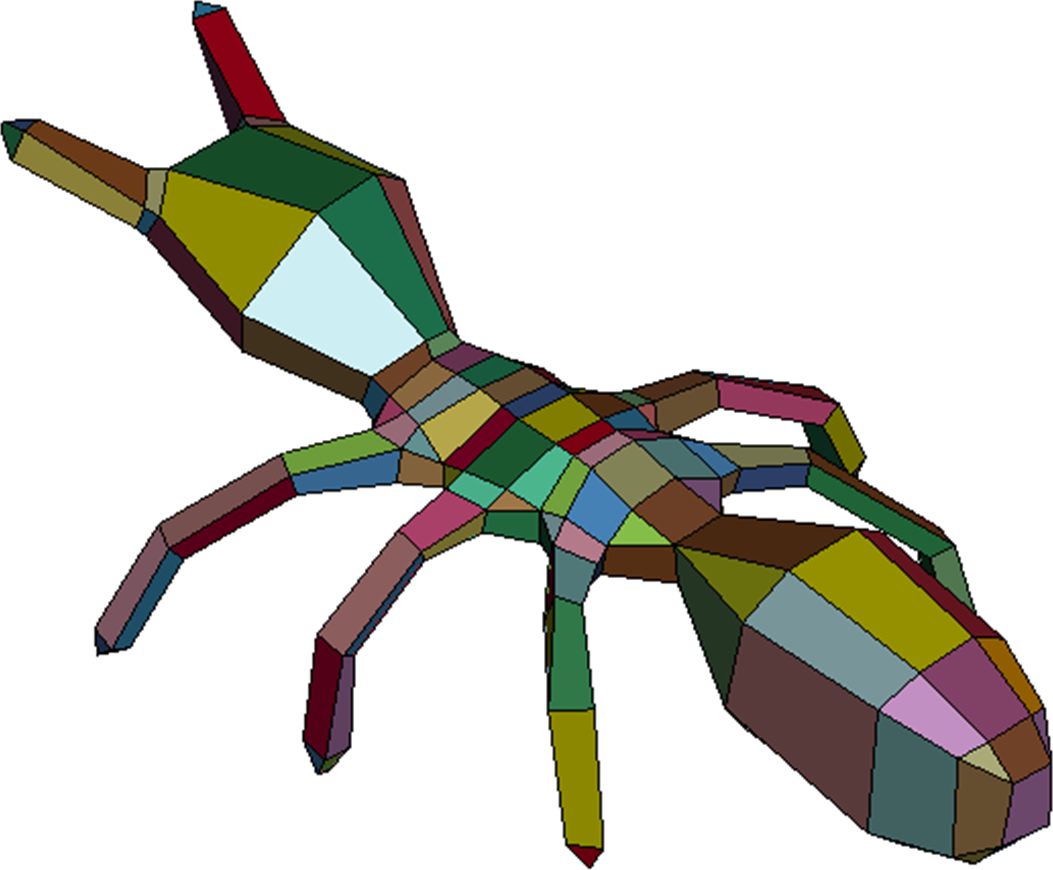}
&\includegraphics[width=0.33\textwidth]{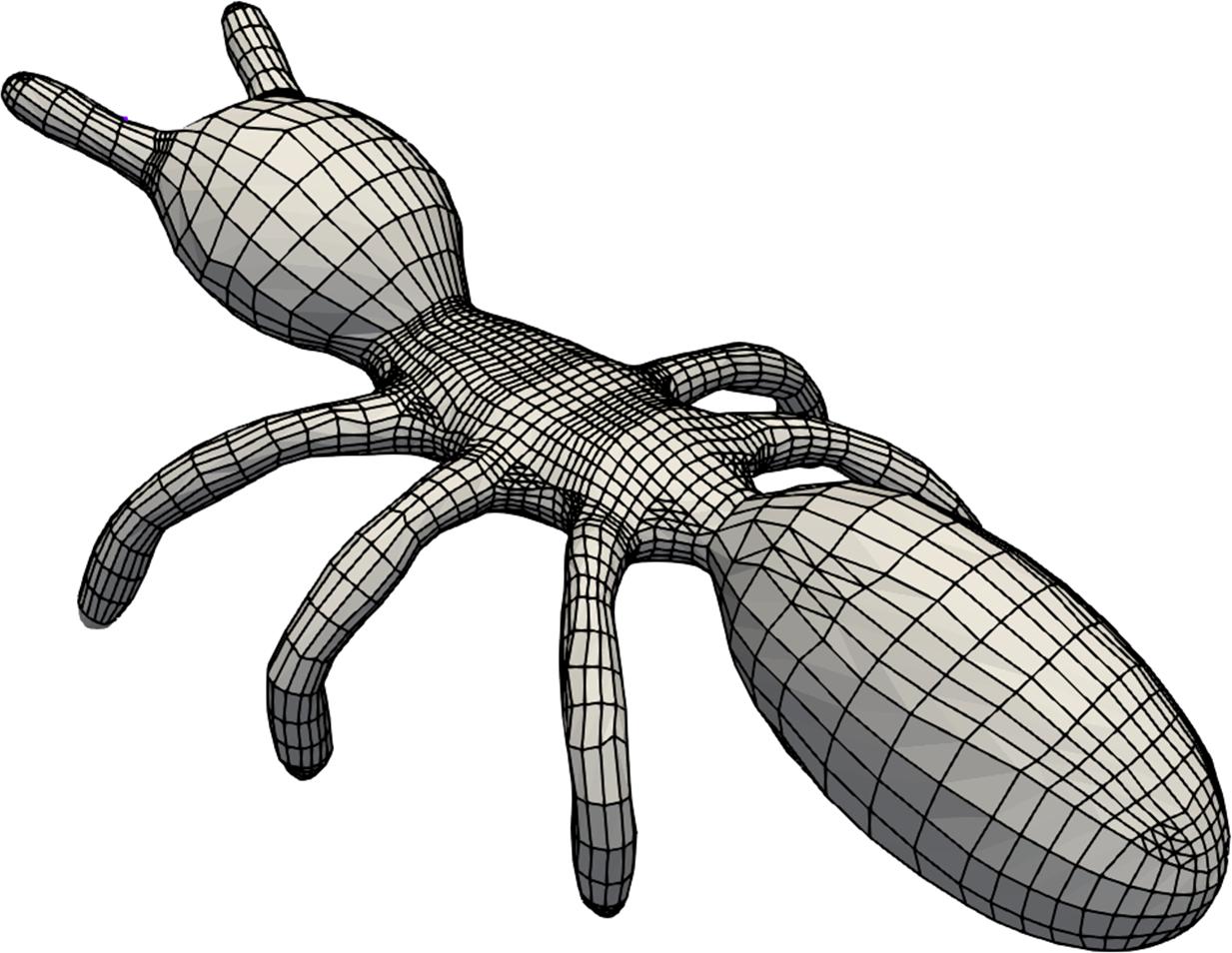}\\
  \includegraphics[height=0.3\linewidth]{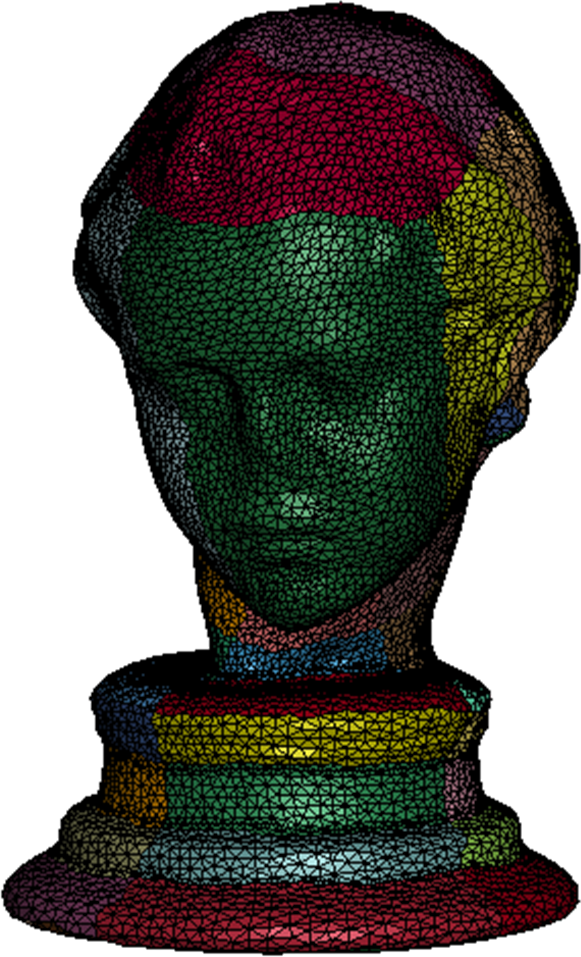}
&\includegraphics[height=0.3\linewidth]{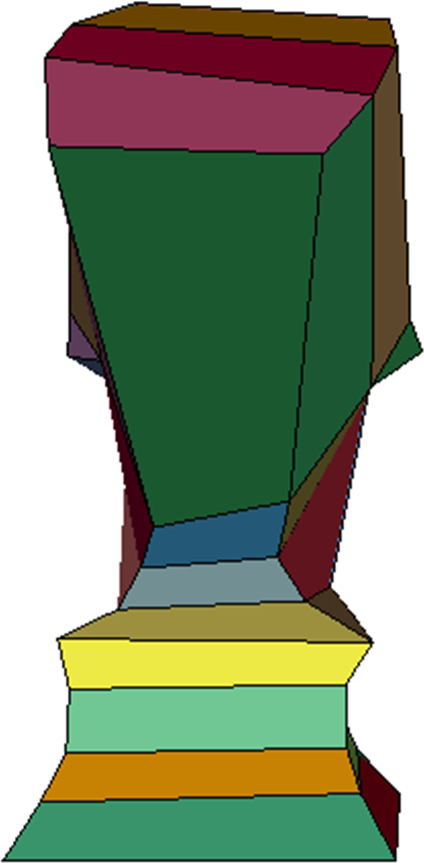}
&\includegraphics[height=0.3\linewidth]{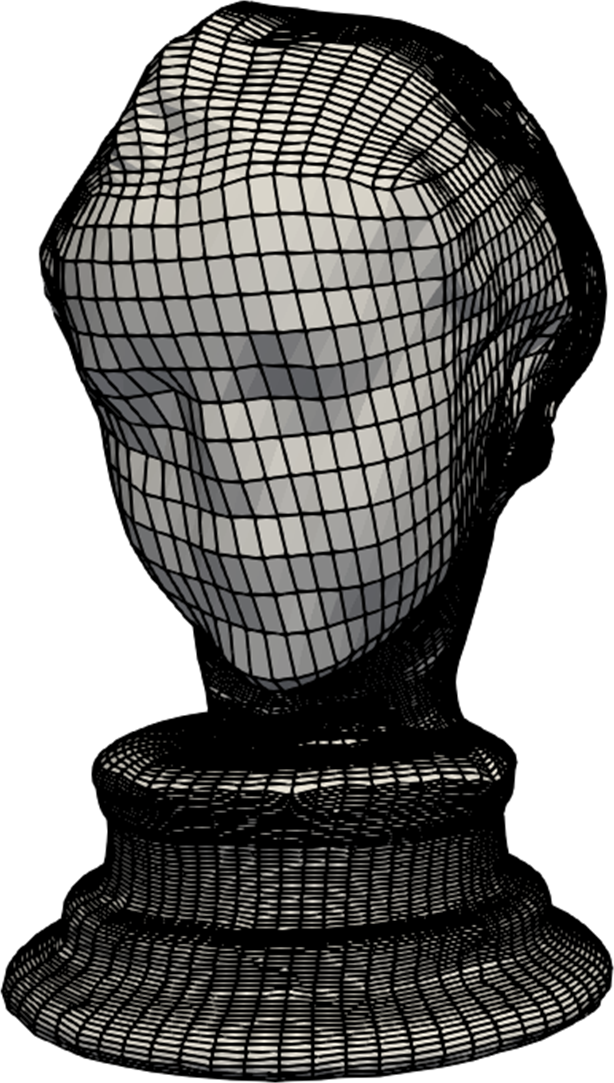}\\
\includegraphics[height=0.3\linewidth]{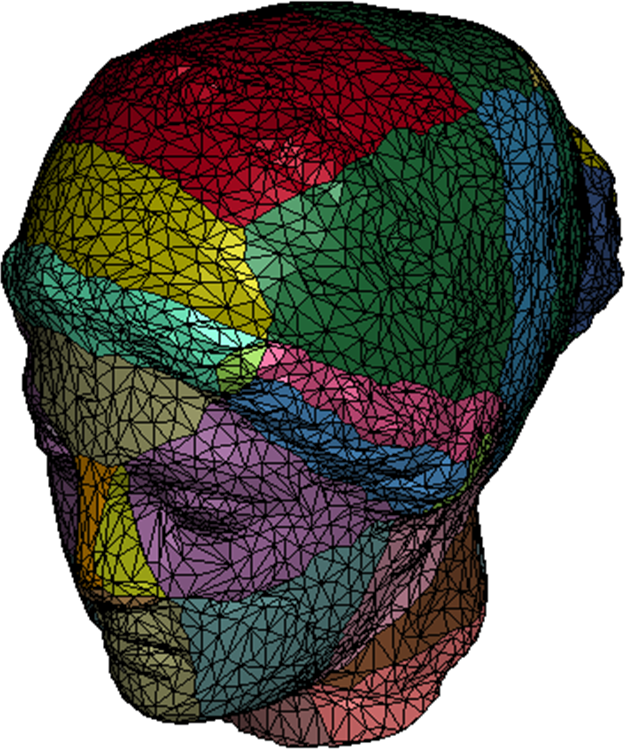}
&\includegraphics[height=0.3\linewidth]{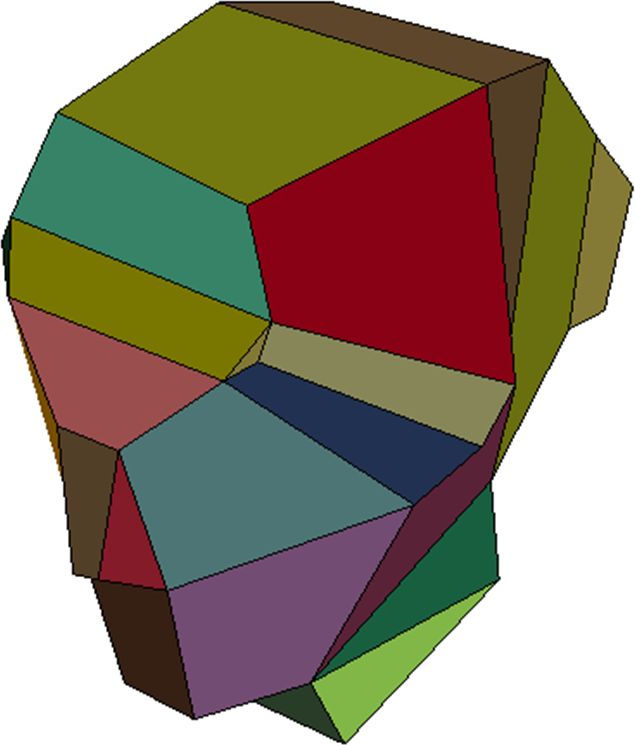}
&\includegraphics[height=0.3\linewidth]{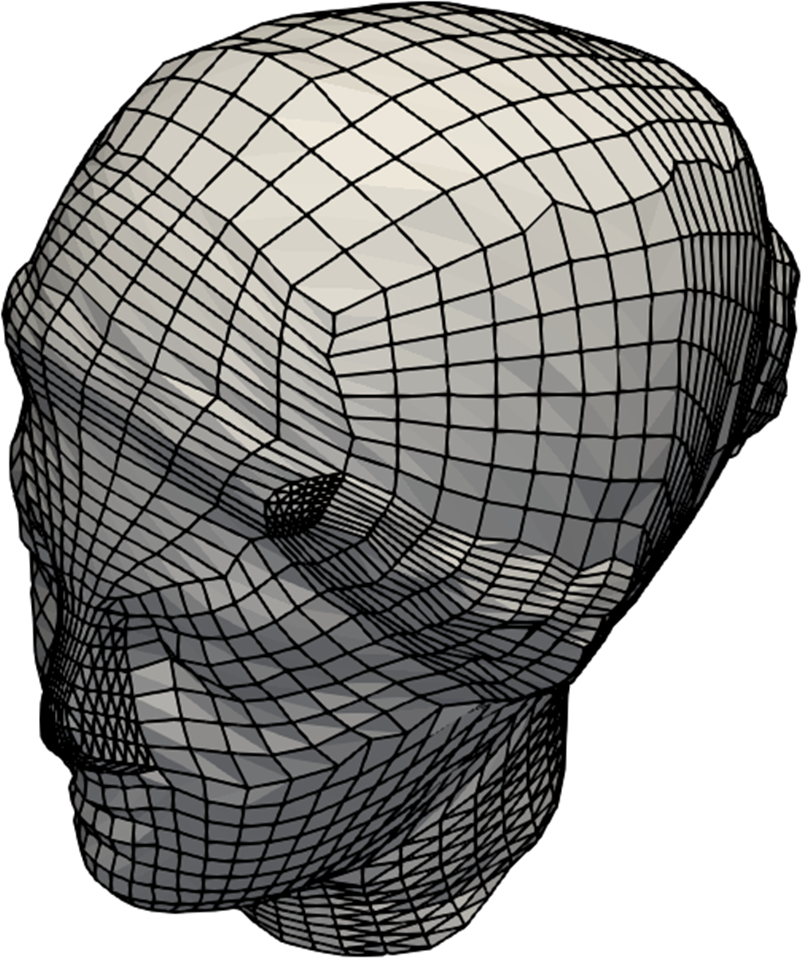}\\
\includegraphics[height=0.3\linewidth]{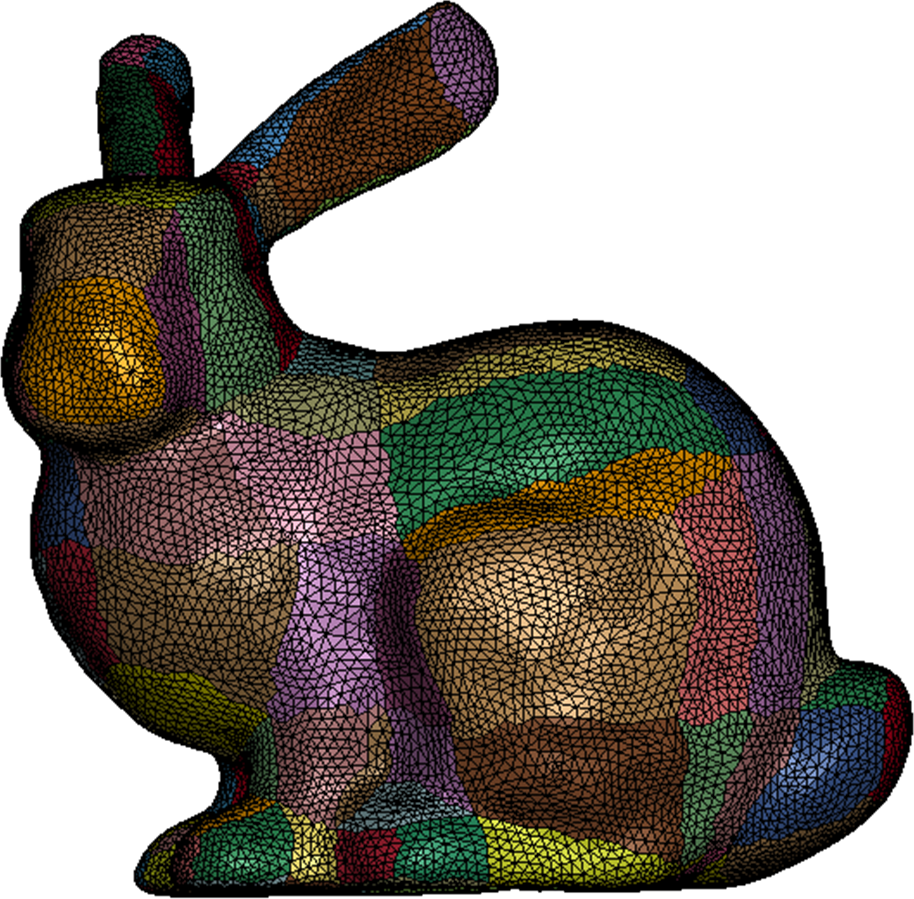}
&\includegraphics[height=0.3\linewidth]{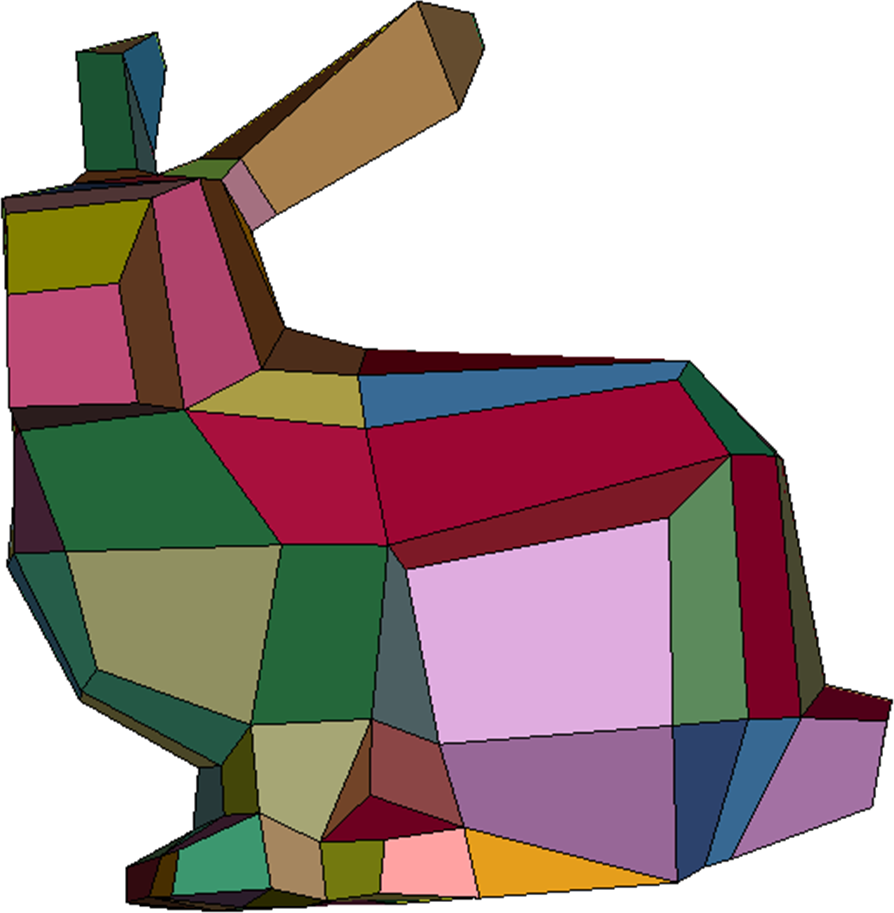}
&\includegraphics[height=0.3\linewidth]{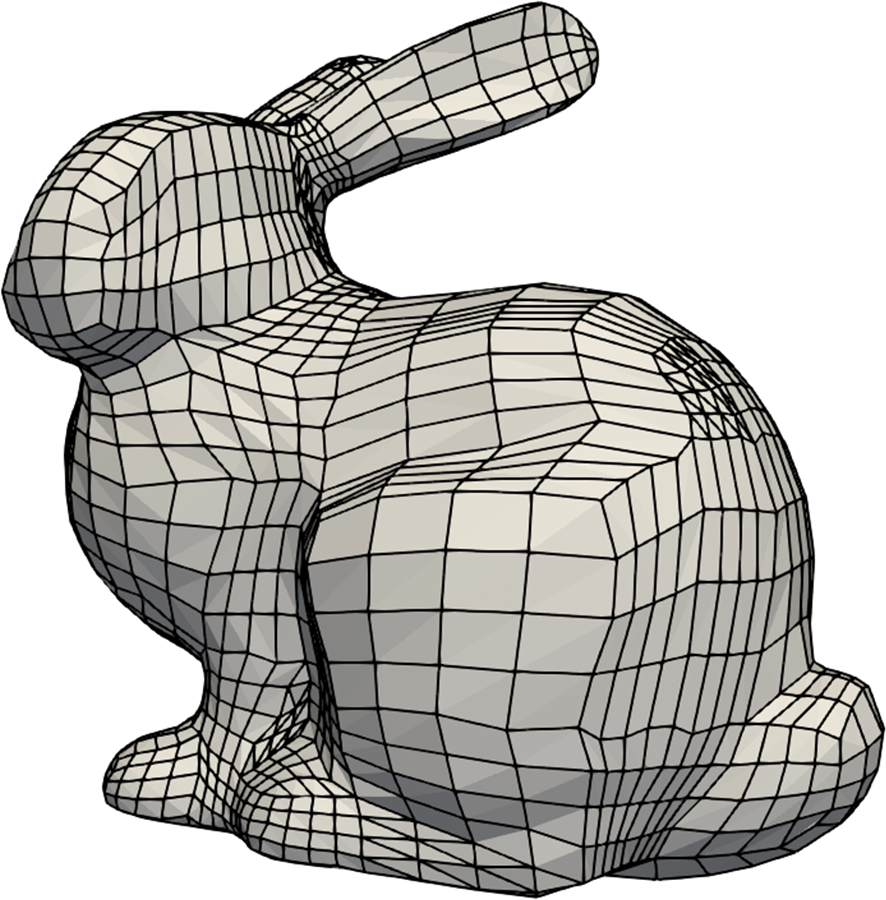}\\
   (a) & (b) & (c)\\
\end{tabular}
\caption{ Results of fertility, ant, bust, igea, and bunny models. (a) Surface triangle meshes and
  segmentation results; (b) polycube structures; and (c) hex-dominant
   meshes. }
    \label{fig:model2}
  \end{figure}

We have applied the software package to several models and generated
hex-dominant meshes with good quality. For each model, we show the
segmentation result, the corresponding polycube structure, and the
hex-dominant mesh. These models include: rockerarm
(Fig.~\ref{fig:Pipeline_Structure}); two types of mount, hepta and a
base with four holes (Fig.~\ref{fig:model1}); fertility, ant, bust,
igea, and bunny (Fig.~\ref{fig:model2}).
Table~\ref{Polycube_Table_1} shows the statistics of all tested
models.  We use the scaled Jacobian to evaluate the quality of hex
elements. The aspect ratio is used as the mesh quality metric for
prism and tet elements which is the ratio between the longest and
shortest edges of an element. The aspect ratio is computed with the
LS-PrePost, which is a pre and post-processor for
LS-DYNA~\cite{ls2007manual}. From Table~\ref{Polycube_Table_1}, we can
observe that the generated hex-dominant meshes have good quality
(minimal Jacobian of hex elements
$>0.1$). Figs.~\ref{fig:model1}-\ref{fig:model2}(a) show the
segmentation results of the testing models. Then, we generate
polycubes (Figs.~\ref{fig:model1}-\ref{fig:model2}(b)) based on the
surface segmentation. Finally, we generate hex-dominant meshes
(Figs.~\ref{fig:model1}-\ref{fig:model2}(c)).

\begin{table}[htp]
\caption{Statistics of all the tested models.}
\label{Polycube_Table_1}
\centering
\setlength{\tabcolsep}{3.7pt}
\scriptsize
\begin{tabular}{|c|c|c|c|c|c|c|c|}
  \hline
  Model &Input triangle mesh  &\multicolumn{3}{c|}{Number of elements}&
                                                                        Jacobian (worst)&\multicolumn{2}{c|}{Aspect
                                                                                          ratio
                                                                                          (min, max)}\\
  \cline{3-5} \cline{7-8}
        &(vertices, elements)& Hex&Prism&Tet&
                                              Hex&Prism&Tet\\
  \hline
  rockerarm (Fig.~\ref{fig:Pipeline_Structure})&(11,705,
                                                 23,410)&3,840&704&
                                                                    128&0.20&(1.32,
                                                                              4.93)&
                                                                                     (1.62,
                                                                                     2.98)\\
  mount1 (Fig.~\ref{fig:model1})&(929, 1,868)&4,224&640&
                                                         128&0.20&(1,91,
                                                                    19.57)&
                                                                            (1.67, 3.69)\\
  mount2 (Fig.~\ref{fig:model1})&(1,042, 2,096)&6,720&1,024&
                                                 128&0.28&(2.63,
                                                            9.54)&
                                                                   (2.30, 4.03)\\
  hepta (Fig.~\ref{fig:model1})&(692, 1,380)&3,776& 1,280&
                                              128&0.50&(1.51, 4.48)&
                                                                     (1.63, 3.14)\\
  base  (Fig.~\ref{fig:model1})&(5,342, 10,700)&3,712& 384&
                                                 128&0.34&(1.28, 2.33)& (1.70,
                                                                         8.31)\\
  fertility (Fig.~\ref{fig:model2})&(6,644, 13,300)&2,752& 320&
                                                     128&0.20&(2.96, 11.40)&
                                                                              (1.69,
                                                                              2.69)\\
  ant (Fig.~\ref{fig:model2})&(7,309, 14,614)&4,480& 1,536&
                                               128&0.21&(1.16, 6.52)& (1.60, 2.79)\\
  bust (Fig.~\ref{fig:model2})&(12,683, 25,362)&118,272& 20,480&
                                                                 8,192&0.10&(1.35, 48.66)& (1.56,
                                                                           4.96)\\
  igea (Fig.~\ref{fig:model2})&(4,532, 9,060)&6,016& 3,584&
                                               1,024&0.21&(1.61, 12.25)&(1.58, 4.69)\\
  bunny (Fig.~\ref{fig:model2})&(14,131, 28,258)&2,752& 1,472&
                                                  128&0.20&(1.44, 10.08)& (1.98,
                                                                           4.18)\\
  \hline
\end{tabular}
\end{table}

\vspace*{-9mm}
\section{Conclusion and future work}

In this paper, we present a new HexDom software package to generate
hex-dominant meshes. The main goal of HexDom is to extend the
polycube-based method to hex-dominant mesh generation. The compiled
software package makes our pipeline accessible to industrial and
academic communities for real-world engineering applications. It
consists of six executable files, namely segmentation module
(Segmentation.exe), polycube construction module (Polycube.exe),
hex-dominant mesh generation module (HexGen.exe, PrismGen.exe,
TetGen.exe) and quality improvement module (Quality.exe). These
executable files can be easily run in the Command Prompt platform. The
rockerarm model was used to explain how to run these programs in
detail. We also tested our software package using several other
models.

Our software has limitations which we will address in our future
work. First, the hex-dominant mesh generation module is semi-automatic
and needs user intervention to create polycube structure. Second, the
degenerated cubic regions and non-degenerated cubic regions need to
be handled separately. We will improve the underneath algorithm and
make polycube construction more automatic. In addition, we will also
develop spline basis functions for tetrahedral and prism elements to
support isogeometric analysis for hybrid meshes.
\vspace*{-7mm}
\section*{Acknowledgment}
Y. Yu, J. Liu and Y. Zhang were supported in part by Honda funds. We
also acknowledge the open source scientific library Eigen and its
developers.
\vspace*{-7mm}
\bibliography{Hex_dominant_hex} \bibliographystyle{spmpsci}
\end{document}